\documentclass[10pt,aps,prb,twocolumn, nofootinbib]{revtex4-2}
\usepackage{amssymb,amsmath}
\usepackage{subfig}
\usepackage{cases}
\usepackage{slashed}
\usepackage{dsfont}
\usepackage{mathtools}
\usepackage{graphicx}
\usepackage{color}
\usepackage{braket}
\usepackage{commath}
\usepackage{hyperref}
\usepackage{listings}
\hypersetup{
    unicode=true,          % non-Latin characters in Acrobat’s bookmarks
    pdftoolbar=true,        % show Acrobat’s toolbar?
    pdfmenubar=true,        % show Acrobat’s menu?
    pdffitwindow=false,     % window fit to page when opened
    pdfstartview={FitH},    % fits the width of the page to the window
    pdfauthor={Kevin Multani},     % author
    colorlinks=true,       % false: boxed links; true: colored links
    linkcolor=blue,          % color of internal links
    citecolor=red,        % color of links to bibliography
}
\graphicspath{{figures/}}
\begin{document}
\preprint{APS/123-QED}
\title{Formal Green's-function theory in non-Hermitian lattice systems}
\author{Changrui Chen}
\affiliation{Department of Physics, Beijing Normal University, Beijing 100875, China}
\author{Wenan Guo}
\email{waguo@bnu.edu.cn}
\affiliation{Department of Physics, Beijing Normal University, Beijing 100875, China}
\affiliation{Key Laboratory of Multiscale Spin Physics (Ministry of Education), Beijing Normal University, Beijing 100875, China}
\date{May 6, 2024}

\begin{abstract}
In this paper, we employ the generalized Bloch theory to rediscover the generalized Brillouin zone theory and follow this way to obtain Green's function of the non-Hermitian system. We focus on a classical chiral model and give the exact expression of the Green function for a finite-size system and the formal expression of the Green function suitable for infinite size. Based on these results, we further derive the correlation matrix
%using our theoretical framework 
and validate it numerically against direct calculations for a system of size 40. The numerical results show the accuracy of our exact expression and the high fidelity of our formal expression. 
\end{abstract}

\maketitle
\section{Introduction}
Non-Hermitian systems have been a very active research topic in recent years, which have many interesting properties different from Hermitian systems, such as skin effect and exceptional points (EPs) \cite{doi:10.1080/00018732.2021.1876991}. The most successful theories describing non-Hermitian systems are the generalized Brillouin Zone (GBZ) theory and its relative theories \cite{PhysRevLett.121.086803, PhysRevLett.123.066404, PhysRevLett.125.226402,2021}. The GBZ theory not only precisely explains the transitional points of the topological phase\cite{PhysRevLett.121.086803} but also recovers the breakdown of the conventional Bloch-band picture and bulk-boundary correspondence in non-Hermitian systems \cite{PhysRevLett.123.066404, PhysRevLett.125.226402}. Beyond these, the GBZ theory finds practical applications in calculating energy spectra,  determining winding numbers \cite{PhysRevLett.121.086803}, and extends its influence into diverse areas such as wave dynamics and chiral damping \cite{2021}. In the original paper\cite{PhysRevLett.121.086803} addressing the GBZ theory, Yao and Wang point out the necessity of extending the Brillouin Zone to the complex plane in non-Hermitian systems. This extension is inspired by a similarity transformation, wherein the non-Hermitian Hamiltonian is transformed into a Hermitian one. Following a prescribed ansatz, they derive the generalized Brillouin Zone (GBZ).
%In this paper, we adopt an alternative approach, namely the generalized Bloch theory \cite{PhysRevB.96.195133, PhysRevB.98.245423}, to rediscover the GBZ theory and follow this road to investigate Green's function of the system. 

Green's function stands as a powerful method widely employed in condensed matter physics, serving as a valuable tool for elucidating a system's response to perturbation and capturing its dynamic behavior. In the realm of non-Hermitian physics, recent reports, such as \cite{PhysRevLett.124.056802}, highlight the utility of Green's function in classifying boundary modes and discerning topological properties. 
%With the help of the thus found Green's function, we derive the correlation matrix, which in turn allows the exploration of entanglement spectra and the system's topological properties. 
The Green's function of non-Hermitian systems has been studied previously \cite{hu2023greens, PhysRevB.103.L241408, PhysRevB.107, PhysRevB.105.045122}. In \cite{hu2023greens, PhysRevB.103.L241408}, the authors present a simple integration expression for the Green's function in the bulk regime using matrix-valued Laurent polynomials. However, they leave out the expression for the boundary regime. A similar expression is given in \cite{PhysRevB.107}, where they take the approach by constructing minimally biorthogonal bases for the deep bulk but don't discuss the edge components. Another work, \cite{PhysRevB.105.045122}, gives the exact summation expression for a non-Hermitian system from end to end but skips details about the bulk regime. 

In this paper, we adopt an alternative approach to the way utilized by Yao and Wang \cite{PhysRevLett.121.086803}, namely the generalized Bloch theory \cite{PhysRevB.96.195133, PhysRevB.98.245423}, to rediscover the GBZ theory. %and follow this road to
This also allows us to derive the exact Green's function of the system. 
%including both bulk and boundary parts. 
Comparing to \cite{hu2023greens, PhysRevB.103.L241408, PhysRevB.107, PhysRevB.105.045122}, our formula not only offers an exact summation expression tailored for finite sizes but also provides a formal expression in integral form suitable for infinite sizes. Importantly, our theoretical framework covers both the boundary and bulk areas, with results matching well with numerical outcomes for both exact and formal expressions. 
%comparing to previous results \cite{hu2023greens, PhysRevB.103.L241408, PhysRevB.107, PhysRevB.105.045122}.
Our present work is reduced to 
the formalisms presented in \cite{hu2023greens, PhysRevB.103.L241408, PhysRevB.107}, if we ignore the boundary part and the edge part of Green's function, as presented by $G_{bound}$ and $G_{edge}$ in Eq. \eqref{ee40}, despite debates surrounding the formalism of the proposed Green's function in \cite{Meden_2023}.
With the help of the thus found Green's function, we derive the correlation matrix, which in turn allows the exploration of entanglement spectra and the system's topological properties. 

The paper is organized as follows: In Sec. \ref{sec:one}, we introduce the generalized Bloch theory in detail, and in Sec. \ref{model}, we use this theory to obtain the wavefunction of the non-Hermitian SSH model, which is our primary concern in this paper. Sec. \ref{sec:4} details how we obtain the exact expression of the Green function of a non-Hermitian system in finite size and the formal expression as the size tends to infinity. Finally, in Sec. \ref{nu}, we verify our discovery numerically, and at last, we summarize the paper.
\section{\label{sec:one}The Generalized Bloch Theory }

Let's start by summarizing the generalized Bloch theory \cite{PhysRevB.96.195133}. Consider a translational invariant Hamiltonian $H$ in one-dimension despite of its hermiticity, i.e. $\bra{i}H\ket{j}=\bra{i+1}H\ket{j+1}$ for arbitrary $\ket{i}$ and $\ket{j}$ which denote the lattice site. Such Hamiltonian $H$ has following properties: $\bra{i}H\ket{j}=h_{j-i}$, $h_l$ is a $n\times n$ matrix, if there are $n$ types of particles in one unit cell. For $n=1$, $H$ is referred to as a Toeplitz matrix under lattice representation. For $n>1$, $H$ is a block Toeplitz matrix \cite{CIT-006,hirschman1965studies,Cobanera_2017}. If $h_l=0$ for any $|l|>R$, where $R$ is an integer, $H$ is termed a banded (block) Toeplitz matrix.

It is imperative to note that, currently, the length of the 1D chain $N$ or the range of $i$ and $j$, and the boundary conditions remain unspecified. In the context of a banded Toeplitz matrix $H$, assuming a sufficiently extensive chain ($N > 2R$), the equation:
\begin{equation}\label{fi}
    H \ket{\varepsilon}=\varepsilon\ket{\varepsilon},
\end{equation}
is equivalent to the following two conditions:
\begin{subnumcases}{}
P_B H \ket{\varepsilon}=\varepsilon P_B\ket{\varepsilon}\label{b};\\
P_{\partial}H \ket{\varepsilon}=\varepsilon P_{\partial}\ket{\varepsilon}\label{p}.
\end{subnumcases}
Here, $P_B = \sum_{j=R+1}^{N-R}\ket{j}\bra{j}$, $P_{\partial}=\mathds{1}-P_B=\sum_{j=1}^{R}\ket{j}\bra{j}+\sum_{j=N-R+1}^{N}\ket{j}\bra{j}$. The integer $R$ represents the boundary or hopping range. Eq. \eqref{p} contains boundary information, while Eq. \eqref{b} is the bulk equation. 

To delve further into Eq. \eqref{b}. let's examine an infinite Toeplitz matrix $H_{\infty}$, where $\bra{i}H_{\infty}\ket{j}=h_{j-i}$, for $i,j\in\mathds{Z}$. If $\ket{\psi}$ satisfies: 
\begin{equation}\label{if}
    H_{\infty}\ket{\psi}=\varepsilon\ket{\psi},
\end{equation}
its projection onto the finite lattice $P_{1,N}\ket{\psi}$ must satisfies Eq. \eqref{b}, where $P_{1,N}=\sum_{i=1}^{N}\ket{i}\bra{i}$. Therefore, solving Eq. \eqref{b} suffices. As detailed  in \cite{PhysRevB.96.195133}, the translational operator $T=\sum_{j\in\mathds{Z}}\ket{j}\bra{j+1}$ is commutated with $H_{\infty}$, $([T,H_{\infty}]=0)$. Thus, it's eigenvectors $\{\ket{z}=\sum_{j\in\mathds{Z}}z^j\ket{j}, z\in\mathds{C}\}$ are also the eigenvectors of $H_{\infty}$. We have:
\begin{align}
    T\ket{z}&=z\ket{z};\label{4}\\
    H_{\infty}\ket{u(z)}\ket{z}=h(z)&\ket{u(z)}\ket{z}=\varepsilon\ket{u(z)}\ket{z}.\\
    \text{Where}\quad h(z)=\sum_{l=-R}^{R}h_lz^l,&\quad h(z)\ket{u(z)}=\varepsilon\ket{u(z)}.\label{bulk}
\end{align}
The deriving of the above equations can also be found in \cite{PhysRevB.96.195133}. 

Note that $h(z)$ is commonly referred to as the bulk Hamiltonian \cite{Asboth2016,2021}. To solve Eq. \eqref{b}, we only need to solve Eq. \eqref{bulk}, a $n\times n$ matrix equation. For a given $\varepsilon$, the corresponding $z$ is not unique. Considering the characteristic equation $det(h(z)-\varepsilon\mathds{1})=0$, it can have at most $2R n$ different roots for a given $\varepsilon$. Different $z$ corresponds to different $\ket{u(z)}$.Therefore, for arbitrary $\{c_k\}\in \mathds{C}$, 
\begin{equation}\label{psi}
\ket{\psi}=\sum_kc_k\ket{z_k}\ket{u(z_k)},
\end{equation}
is the solution of Eq. \eqref{b} when projected onto the finite lattice. Where $z_k$ represents different roots of the characteristic equation\footnote{For the case involving multiple roots, refer to the details in \cite{PhysRevB.96.195133}.}. To obtain the final solution of Eq. \eqref{fi}, we substitute Eq. \eqref{psi} into Eq. \eqref{p} to determine which $\{z_k\}$s are allowed and the value of $\{c_k\}$s. In other words, we look for a non-zero solution for the equation:
\begin{equation}\label{constraint}
    B(\varepsilon)\begin{pmatrix}
        c_1\\\vdots\\c_k\\\vdots
    \end{pmatrix}=0,
\end{equation}
where
\begin{equation}
    B({\varepsilon})=\begin{pmatrix}
        P_{\partial}(H-\varepsilon\mathds{1})\ket{u(z_1)}\ket{z_1}\\\vdots\\P_{\partial}(H-\varepsilon\mathds{1})\ket{u(z_k)}\ket{z_k}\\\vdots
    \end{pmatrix}^{\mathrm{T}}.
\end{equation}
$B(\varepsilon)$ is also known as the boundary matrix. The constraint Eq. \eqref{constraint} has non-zero solutions determine the energy spectrum for $H$.

The procedure outlined above to solve the system under any boundary condition is known as the generalized Bloch Theory, and its detailed expression can be found in reference \cite{PhysRevB.96.195133}. It's noteworthy that in this procedure, the hermiticity of $H$ is not considered, indicating the applicability of this method to non-Hermitian systems.

\section{Model\label{model}}
\begin{figure}[h]
\centering
\includegraphics[width=1.0\linewidth]{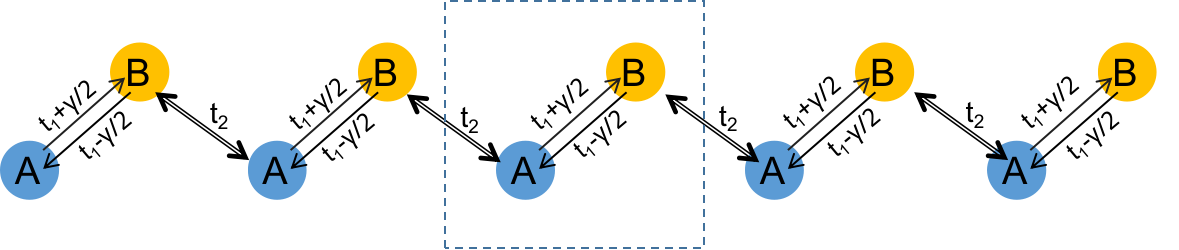}
\caption{Non-Hermitian SSH model. The dotted box indicates a unit cell.}
\label{fig1}
\end{figure}

In this section, we employ the generalized Bloch theory to investigate the non-Hermitian SSH model, as depicted in Fig. \ref{fig1}. 

The chain consists of two sublattices, $A$ and $B$, in each unit cell. The inter-cell transition amplitude is denoted as $t_2$, while the inner-cell transition amplitude exhibits asymmetry. Specifically, the transition from $A$ to $B$ is $t_1 + \gamma/2$, and from $B$ to $A$ is $t_1 - \gamma/2$. The Hamiltonian $H$ for this model is expressed as:
\begin{small}
\begin{equation}\label{H10}
\begin{split}
   H=&\sum_{j}(t_1+\gamma/2)\ket{A,j}\bra{B,j}+(t_1-\gamma/2)\ket{B,j}\bra{A,j}\\+&
   \sum_{j}t_2\left(\ket{A,j+1}\bra{B,j}+\ket{B,j}\bra{A,j+1}\right).
\end{split}
\end{equation}
\end{small}

This Hamiltonian manifests as a banded block Toeplitz matrix with specific blocks $h_l$ given by:
\begin{equation}\label{h01}
\begin{split}
    &h_0=\begin{pmatrix}
        0&t_1+\frac{\gamma}{2}\\t_1-\frac{\gamma}{2} &0
    \end{pmatrix},\\
     &h_1=\begin{pmatrix}
        0 &0\\t_2 &0
    \end{pmatrix},\quad h_{-1}=\begin{pmatrix}
        0&t_2\\0&0
    \end{pmatrix}.
\end{split}    
\end{equation}
while all the other $h_l$ is $0$, which suggests that $R=1$. Its bulk Hamiltonian can be written as:
\begin{equation}\label{13} 
\begin{split}
h(z)=&h_0+h_1z+h_{-1}z^{-1}\\=&\begin{pmatrix}
0&t_1+\frac{\gamma}{2}+t_2z^{-1}\\t_1-\frac{\gamma}{2}+t_2z&0      
    \end{pmatrix}.
\end{split}
\end{equation}
The characteristic equation of this bulk Hamiltonian is expressed as:
\begin{align}\label{eq13}
    &det(h(z)-\varepsilon\mathds{1})=0\\ \Leftrightarrow&(t_1+\frac{\gamma}{2})t_2z^2+(t_1^2-\frac{\gamma^2}{4}+t_2^2-\varepsilon^2)z+(t_1-\frac{\gamma}{2})t_2=0 \nonumber.\end{align}
This quadratic equation implies two $z$ values that satisfy the equation for a specific $\varepsilon$ in the norm sense. Denoting these roots as $z$ and $z'$, we consider the ansatz:
\begin{equation}\label{con}
    \ket{\psi_{\varepsilon}}=c_z\ket{u(z)}\ket{z}+c_{z'}\ket{u(z')}\ket{z'},
\end{equation}
where
\begin{equation}\label{open}
     h(z)\ket{u(z)}=\varepsilon\ket{u(z)}, \quad
     h(z')\ket{u(z')}=\varepsilon\ket{u(z')}.
\end{equation}
The impact of boundary conditions on the system will be explored next.

\subsection{Periodic Boundary Condition}
The Hamiltonian presented in Eq. \eqref{H10} is expressed in a casual form, lacking information about the boundary in the sum. For a periodic boundary, the appropriate representation is:
\begin{small}
\begin{equation}\label{H16}
\begin{split}
   H=&\sum_{j=1}^{N}(t_1+\gamma/2)\ket{A,j}\bra{B,j}+(t_1-\gamma/2)\ket{B,j}\bra{A,j}\\+&
   \sum_{j=1}^{N-1}t_2\left(\ket{A,j+1}\bra{B,j}+\ket{B,j}\bra{A,j+1}\right)\\+&t_2(\ket{A,1}\bra{B,N}+\ket{B,N}\bra{A,1}).
\end{split}
\end{equation}
\end{small}
Given that $R=1$, the boundary projection operator is $P_{\partial}=\ket{1}\bra{1}+\ket{N}\bra{N}$. Utilizing Eq. \eqref{13}, we can derive $B(\varepsilon)$ as follows\footnote{\label{foot2}To obtain Eq. \eqref{ee17}, you can try to rewrite Eq. \eqref{H16} as:
\begin{equation}
    H=\begin{pmatrix}
        h_0&h_1&\cdots&h_{-1}\\
        h_{-1}&h_0&h_1&\cdots\\
        \vdots&\vdots&\ddots&\vdots\\
        h_1&\cdots&h_{-1}&h_0
    \end{pmatrix}.
    \nonumber
\end{equation}
And use properties Eq. \eqref{13} and Eq. \eqref{open}}:
\begin{equation}\label{ee17}
    B(\varepsilon)=\begin{pmatrix}
        h_{-1}(z^N-1)\ket{u(z)}&h_{-1}(z'^N-1)\ket{u(z')}\\h_1z(1-z^N)\ket{u(z)}&h_1z'(1-z'^N)\ket{u(z')}
    \end{pmatrix}.
\end{equation}
Note that $B(\varepsilon)$ here is a non-trivial $4\times2$ matrix. To obtain a non-zero solution for $(c_z, c_{z'})$, one possibility is to set $z^N=1$, $c_z=1$, $c_{z'}=0$ (similarly, setting $z'^N=1$ results in $c_z=0$, $c_{z'}=1$).

Hence, the wavefunction for the periodic boundary is given by:
\begin{equation}\label{eq18}
   \ket{ \psi_{\varepsilon}}=\ket{u(z)}\ket{z},\quad  z^N=1.
\end{equation}
In the limit as $N\to\infty$, any $|z|=1$ could satisfy this condition.
\subsection{Open Boundary Condition}
For open boundary conditions, the appropriate expression of the Hamiltonian is:
\begin{small}
\begin{equation}\label{H19}
\begin{split}
   H=&\sum_{j=1}^{N}(t_1+\gamma/2)\ket{A,j}\bra{B,j}+(t_1-\gamma/2)\ket{B,j}\bra{A,j}\\+&
   \sum_{j=1}^{N-1}t_2\left(\ket{A,j+1}\bra{B,j}+\ket{B,j}\bra{A,j+1}\right).
\end{split}
\end{equation}
\end{small}
Using Eq. \eqref{13} again, we can derive\footnote{Try to use the same technique as in footnote\ref{foot2}.}:
\begin{equation}
    B(\varepsilon)=\begin{pmatrix}
        -h_{-1}\ket{u(z)}&-h_{-1}\ket{u(z')}\\-h_1z^{N+1}\ket{u(z)}&-h_1z'^{N+1}\ket{u(z')}
    \end{pmatrix}.
\end{equation}
Due to the structure of $h_1$ and $h_{-1}$ in Eq. \eqref{h01}, having only one nonzero row, the condition for $B(\varepsilon)$ to have non-zero solutions is equivalent to:
\begin{equation}\label{eq20}
\begin{pmatrix}
    -t_2\phi_B&-t_2\phi_B'\\-t_2z^{N+1}\phi_A&-t_2z'^{N+1}\phi_A'
\end{pmatrix},
\end{equation}
where we denote:
\begin{equation}\label{u21}
    \ket{u(z)}=\begin{pmatrix}
        \phi_A\\\phi_B
    \end{pmatrix},\quad
    \ket{u(z')}=\begin{pmatrix}
        \phi_A'\\\phi_B'
    \end{pmatrix}.
\end{equation}
Furthermore, due to Eq. \eqref{open}, we have the relations:
\begin{equation}\label{eq22}
\phi_B=\frac{\phi_A}{\varepsilon}(t_1-\frac{\gamma}{2}+t_2z),\quad\phi_B'=\frac{\phi_A'}{\varepsilon}(t_1-\frac{\gamma}{2}+t_2z').
\end{equation}
Substituting Eq. \eqref{eq22} into Eq. \eqref{eq20} and making the determinant of Eq. \eqref{eq20} equal to 0, we get:
\begin{equation}\label{eq23}
    (t_1-\frac{\gamma}{2}+t_2z)z'^{N+1}=(t_1-\frac{\gamma}{2}+t_2z')z^{N+1}.
\end{equation}
Notably, according to Vieta's theorem and Eq. \eqref{eq13}, we can derive $zz'=(t_1-\gamma/2)/(t_1+\gamma/2)\equiv r$. Substituting it into Eq. \eqref{eq23}, we obtain:
\begin{equation}\label{eq24}
    (t_1-\frac{\gamma}{2})z^{2N+2}+t_2rz^{2N+1}-t_2r^{N+1}z-(t_1-\frac{\gamma}{2})r^{N+1}=0.
\end{equation}
This is a $(2N+2)$th-order polynomial, typically having $2N+2$ roots. However, considering Eq. \eqref{eq23}, we observe that multiple roots exist for $z=z'=\pm\sqrt{r}$, which should be excluded, resulting in a total of $2N$ roots.

Therefore, we can obtain the wavefunction of the system in the open boundary condition:
\begin{equation}\label{eq25}
\begin{split}
	&\ket{\psi_{\varepsilon}}=c_z\ket{u(z)}\ket{z}+c_{z'}\ket{u(z')}\ket{z'},\\ &\text{z is the solution of Eq. \eqref{eq24}},z\neq\pm\sqrt{r},z'=r/z.
\end{split}    
\end{equation}
The coefficients $c_z$ and $c_{z'}$ can be readily determined from $B(\varepsilon)$ or Eq. \eqref{eq20}: 
\begin{equation}\label{ab26}
    \frac{c_z}{c_{z'}}=-\frac{\phi_B'}{\phi_B}\quad\text{or}\quad\frac{c_z}{c_{z'}}=-\frac{z'^{N+1}\phi_A'}{z^{N+1}\phi_A}.
\end{equation}
 Notably, these two expressions are equivalent. We will examine the behavior of $z$ as $N$ approaches infinity next.

\begin{figure}[h]
\centering
\includegraphics[width=1.0\linewidth]{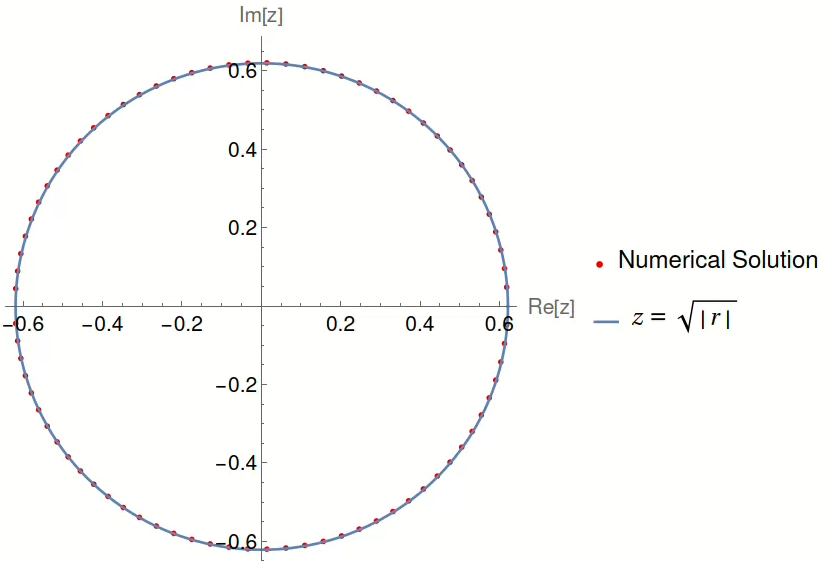}
\caption{The numerical solution of Eq. \eqref{eq24}, when $t_1=3/2,t_2=1,\gamma=4/3$,$N=40$. The blue line represents the circle $|z|=\sqrt{|r|}$ in the complex plane.}
\label{fig2}
\end{figure}

Considering a specific example of a non-Hermitian system with $t_1=3/2$, $t_2=1$, $\gamma=4/3$, and $N=40$. Solving Eq. \eqref{eq24} for this example using Mathematica, we plot the solutions in the complex plane, as depicted in Fig. \ref{fig2}.

We observe that all solutions of Eq. \eqref{eq24} lie on the circle $|z|=\sqrt{|r|}$. This outcome aligns with the findings in Ref. \cite{PhysRevLett.121.086803}, where it is stated that, as $N\to\infty$, $z$ and $z'$ in Eq. \eqref{eq23} must share the same absolute value. Otherwise, one side of Eq. \eqref{eq23} tends toward zero as $N$ grows. Consequently, in the limit of $N\to\infty$, any $z$ satisfying $|z|=\sqrt{|r|}$ can generate a wavefunction in Eq. \eqref{eq25}\footnote{Except $z=\pm\sqrt{r}$, but these seem unimportant in the context of $N\to\infty$.}.

In Ref. \cite{2021,PhysRevLett.121.086803}, $|z|=\sqrt{|r|}$ is termed the generalized Brillouin Zone (GBZ), distinguishing it from the conventional Brillouin Zone in Hermitian systems. Employing the techniques of generalized Bloch theory, we deduce the GBZ for the non-Hermitian system under both periodic and open boundary conditions. Our analysis reveals that the GBZ is contingent on the boundary condition—specifically, $|z|=\sqrt{|r|}$ for open boundary conditions and $|z|=1$ for periodic boundary conditions.
\section{The Green's function of non-Hermitian system\label{sec:4}}
One of the main objectives of our article is to derive the Green's function for the non-Hermitian system. For a non-Hermitian system with wave functions denoted as $\{\ket{\psi_{\varepsilon}}\}$, its Green's function is expressed as follows:
\begin{equation}\label{G}
    G(\omega)=\sum_{\varepsilon}\frac{\ket{\psi_{\varepsilon}}\bra{\Tilde{\psi_{\varepsilon}}}}{\braket{\Tilde{\psi_{\varepsilon}}|\psi_{\varepsilon}}}\frac{1}{\omega-\varepsilon},
\end{equation}
where $H\ket{\psi_{\varepsilon}}=\varepsilon\ket{\psi_{\varepsilon}}$ and $\bra{\tilde{\psi}_{\varepsilon}}H=\varepsilon\bra{\tilde{\psi}_{\varepsilon}}$. 
For the periodic boundary condition, substituting Eq. \eqref{eq18} into Eq. \eqref{G} yields:
\begin{equation}\label{eq37}
    G(\omega)=\sum_l\sum_z\frac{\ket{u_l(z)}\bra{\tilde{u}_l(z)}}{N(\omega-\varepsilon_l(z))}\ket{z}\bra{\bar{z}^{-1}},
\end{equation}
where $\bra{\tilde{u}(z)}$ is the left eigenvector of $h(z)$. The detailed deduction is provided in Appendix \ref{A}.

It's important to note that we have cleverly replaced the sum over $\varepsilon$ with a sum over $z$. Because in Eq. \eqref{eq13}, one $z$ may have multiple corresponding $\varepsilon$s, denoted as $\varepsilon_l(z)$, the summation should involve all possible $\varepsilon_l(z)$ and their corresponding eigenvector $\ket{u_l(z)}$ for a specific $z$. In the specific case of the system under consideration, which exhibits chiral symmetry, every $z$ in the GBZ corresponds to two $\varepsilon$s, satisfying $\varepsilon_1(z)=-\varepsilon_2(z)$. However, in a more general scenario, the GBZ may be decomposed into different sub-GBZs, each corresponding to an independent closed curve in the complex plane. Each sub-GBZ corresponds to a Riemann surface of the solution of Eq. \eqref{eq13} \cite{PhysRevLett.125.226402}. In such cases, $\sum_{\varepsilon}\leftrightarrow\sum_l\sum_{z\in GBZ_l}$.

Now, let's explore the case as $N\to\infty$. Assuming that in this limit, $z$s are uniformly distributed in the GBZ, the difference between $z$s becomes:
\begin{equation}
    \Delta z=z(e^{i\frac{2\pi}{N}}-1)\approx \frac{2\pi i z}{N}.
\end{equation}

This leads to the expression:
\begin{equation}\label{eq39}
\begin{split}  
    \bra{k}G(\omega)\ket{j}=&\sum_l\sum_z\frac{\Delta z}{2\pi i z}\frac{\ket{u_l(z)}\bra{\tilde{u}_l(z)}}{\omega-\varepsilon_l(z)}z^{k-j}\\\xrightarrow{N\to\infty}&\sum_l\int_{|z|=1}\frac{dz}{2\pi i z}\frac{\ket{u_l(z)}\bra{\tilde{u}_l(z)}}{\omega-\varepsilon_l(z)}z^{k-j}.
\end{split}
\end{equation}

This expression represents the Green's function of the periodic boundary non-Hermitian system in the limit as $N\to\infty$.

For open boundary conditions, assuming the corresponding left eigenvalue of $\ket{\psi_{\varepsilon}}$ in Eq. \eqref{eq25} is:
\begin{equation}\label{eq40}
    \bra{\tilde{\psi}_{\varepsilon}}=\tilde{c}_{z}\bra{\tilde{u}(z)}\bra{\bar{z}^{-1}}+\tilde{c}_{z'}\bra{\tilde{u}(z')}\bra{\bar{z'}^{-1}}.
\end{equation}

Likewise, denote: 
\begin{equation}\label{u42}
    \bra{\tilde{u}(z)}=\begin{pmatrix}
   \tilde{\phi}_A\\\tilde{\phi}_B    
    \end{pmatrix}^T,\quad\bra{\tilde{u}(z')}=\begin{pmatrix}
   \tilde{\phi}_A'\\\tilde{\phi}_B'
    \end{pmatrix}^T.
\end{equation}

Similar to Eq. \eqref{ab26}, we have:
\begin{equation}\label{eq43}
    \frac{\tilde{c}_z}{\tilde{c}_{z'}}=-\frac{\tilde{\phi}_B'}{\tilde{\phi}_B}\quad\text{or}\quad \frac{\tilde{c}_z}{\tilde{c}_{z'}}=-\frac{z'^{-(N+1)}\tilde{\phi}_A'}{z^{-(N+1)}\tilde{\phi}_A}.
\end{equation}

Since Eq. \eqref{ab26} and Eq. \eqref{eq43} are in ratio form, it's convince to assume $c_z\tilde{c}_z+c_{z'}\tilde{c}_{z'}=1$. Denote $N_{\varepsilon}=\braket{\Tilde{\psi_{\varepsilon}}|\psi_{\varepsilon}}$, substituting Eq. \eqref{eq40} and Eq. \eqref{eq25} into Eq. \eqref{G}, we derive the Green's function:
\begin{equation}\label{eq47}
\begin{split}
    G(\omega)=&\frac{1}{2}\sum_l\sum_{z}\frac{1}{N_{\varepsilon}(\omega-\varepsilon_l(z))}\Big(\ket{u_l(z)}\bra{\Tilde{u}_l(z)}\ket{z}\bra{\bar{z}^{-1}}\\+&2c_{z'}\tilde{c}_{z}\ket{u_l(z')}\bra{\tilde{u}_l(z)}\ket{z'}\bra{\bar{z}^{-1}}\Big).
\end{split}
\end{equation}

The appearance of $\frac{1}{2}$ is because Eq. \eqref{eq25} contains two components $\ket{z}$ and $\ket{z'}$ (See Appendix \ref{A}). Eq. \eqref{eq47} is a crucial result in our paper, precisely delineating the Green's function of a non-Hermitian system under open boundaries for any size. Upon comparison with the periodic boundary case (Eq. \eqref{eq37}), Eq. \eqref{eq47} can be further decomposed into two parts:
\begin{small}
\begin{align}\label{48}
    G_{bulk}(\omega)=&\frac{1}{2}\sum_l\sum_{z}\frac{1}{N_{\varepsilon}(\omega-\varepsilon_l(z))}\ket{u_l(z)}\bra{\Tilde{u}_l(z)}\ket{z}\bra{\bar{z}^{-1}};\nonumber\\
    G_{bound}(\omega)=&\frac{1}{2}\sum_l\sum_{z}\frac{2c_{z'}\tilde{c}_{z}}{N_{\varepsilon}(\omega-\varepsilon_l(z))}\ket{u_l(z')}\bra{\tilde{u}_l(z)}\ket{z'}\bra{\bar{z}^{-1}}.
\end{align}
\end{small}

The nomenclature $G_{bulk}$ for the first part is justified as it closely relates to the Green's function in the periodic boundary, while the second part, named $G_{bound}$, represents an additional term in open boundary conditions. Similar operations are found in Ref. \cite{PhysRevB.103.195157}.

Again we try to understand the case $N\to\infty$. Still assuming that $z$s are located in the GBZ uniformly, since there are $2N$ possible $z$s in open boundary according to Eq. \eqref{eq24}, Eq. \eqref{eq25}, the difference between $z$s becomes:
\begin{equation}\label{p49}
    \Delta z=z(e^{\frac{i\pi}{N}}-1)\approx\frac{\pi i z}{N}.
\end{equation}

In the limit $N\to\infty$, where $N_{\varepsilon}\to N$, the results for $G_{\text{bulk}}(\omega)$ and $G_{\text{bound}}(\omega)$ are given by\footnote{The attentive reader may observe the change in the subscript of the integral in Eq. \eqref{eq48} now spanning over $GBZ_l$. This adjustment doesn't introduce complications in the systems under consideration. One can interpret it as a scenario where two sub-GBZs overlap at $|z|=\sqrt{|r|}$. These equations can be responsible for a more general non-Hermitian system. However, in cases where the sub-GBZs are not circles, one must modify the integration term from
$
    \int_{z\in GBZ_l} \frac{dz}{2\pi i z}$ to $\int_{z\in GBZ_l} \left(\frac{dz}{2\pi i z }-\frac{d|z|}{2\pi i |z| }\right)$. For further details, refer to Appendix \ref{A}.}:
\begin{small}
\begin{align}\label{eq48}
   \bra{k}G_{bulk}(\omega)\ket{j}=&\sum_l\int_{z\in GBZ_l}\frac{dz}{2\pi i z}\frac{\ket{u_l(z)}\bra{\tilde{u}_l(z)}}{\omega-\varepsilon_l(z)}z^{k-j};\nonumber\\
    \bra{k}G_{bound}(\omega)\ket{j}=&\sum_l\int_{z\in GBZ_l}\frac{dz}{2\pi i z}2c_{z'}\tilde{c}_{z}\frac{\ket{u_l(z')}\bra{\tilde{u}_l(z)}}{\omega-\varepsilon_l(z)}z'^kz^{-j}.
\end{align}
\end{small}

 While the deductions presented in Eq. \eqref{eq48} offer a pleasing theoretical framework, certain limitations arise during the transition from finite to infinite systems. Firstly, in the case of open boundaries where $z$s are situated in sub-GBZs with modules not equal to 1, the convergence from Eq. \eqref{48} to Eq. \eqref{eq48} is of the order $|z|^{k-j}/N$. Consequently, when considering the Green's function between the first and last lattice sites, the convergence speed becomes $|z|^N/N$ or $|z|^{-N}/N$. This poses a challenge for convergence as $N\to\infty$ unless $|z|=1$. However, for any fixed $k-j$, regardless of its magnitude, there exists a sufficiently large $N$ allowing Eq. \eqref{48} to converge to Eq. \eqref{eq48}.

Secondly, the coefficient $2c_{z'}\tilde{c}_{z}$ in $G_{\text{bound}}$ is determined by Eq. \eqref{ab26} and Eq. \eqref{eq43}, signifying their dependence not only on $z$ and $z'$ but also on $N$. The ratio $z'^{N+1}/z^{N+1}$, appearing in both Eq. \eqref{ab26} and Eq. \eqref{eq43}, lacks a well-defined limit as $N\to\infty$ from a mathematical perspective. Consequently, the integral of $G_{\text{bound}}$ in Eq. \eqref{eq48} is not well-defined. These limitations render Eq. \eqref{eq48} only a formal theory. Nevertheless, the bulk properties of non-Hermitian systems can still be predicted by $G_{\text{bulk}}$ in Eq. \eqref{eq48}, as supported by our analysis and others \cite{PhysRevB.103.L241408,hu2023greens,PhysRevB.107}.

It is noteworthy that a third part, $G_{\text{edge}}(\omega)$, may emerge due to the presence of edge states. This term is expressed as:
\begin{equation}\label{e39}
G_{\text{edge}}(\omega)=\sum_l\frac{\ket{0_l}\bra{\tilde{0}_l}}{\omega}.
\end{equation}

Here $\ket{0_l}$ and $\bra{\tilde{0}_l}$ represent the $l$th edge state and its corresponding left eigenvector. For the system under consideration, two edge states exist due to chirality. In summary, the full expression for non-Hermitian systems is given by:
\begin{equation}\label{ee40}
G(\omega)=G_{\text{bulk}}(\omega)+G_{\text{bound}}(\omega)+G_{\text{edge}}(\omega).
\end{equation}

The appearance of the third term is contingent on the existence of edge states. While the precise behavior of $G_{\text{bulk}}$ and $G_{\text{bound}}$ when $N\to\infty$ remains an open question, we plan to examine Eq. \eqref{eq48} using numerical methods in the subsequent chapter.
\section{Correlation matrix and The numerical results\label{nu}}
In this chapter, our primary focus is on the correlation matrix of the open boundary non-Hermitian system, a vital tool for describing physical systems with applications ranging from entanglement entropy to various other domains \cite{PhysRevResearch.2.043191, monkman2022entanglement}. We define $\hat{a}^{\dagger}_k$ and $\hat{a}_k$ as the creation and annihilation operator of lattice $\ket{A,k}$; $\hat{b}^{\dagger}_j$ and $\hat{b}_j$ as that of lattice $\ket{B,j}$. $\{\hat{a}_k, \hat{a}^{\dagger}_j\}=\{\hat{b}_k, \hat{b}^{\dagger}_j\}=\delta_{kj}$, $\{\hat{a}_k, \hat{b}^{\dagger}_j\}=\{\hat{b}_k, \hat{a}^{\dagger}_j\}=0$ being fermionic operator, the standard expression for the correlation matrix in the context of non-Hermitian systems is given by Eq. \eqref{53}:
\begin{equation}\label{53}
(C)_{kj}=\bra{\tilde{\Psi}_0}\hat{c}^{\dagger}_k\hat{c}_j\ket{\Psi_0}.
\end{equation}
Here, 
\begin{align}\label{cab}
\hat{c}_j=\hat{a}_j,&\quad\hat{c}_{N+j}=\hat{b}_j,\\ 
\hat{c}^{\dagger}_j=\hat{a}^{\dagger}_j,&\quad\hat{c}^{\dagger}_{N+j}=\hat{b}^{\dagger}_j,\nonumber
\end{align}
for $j=1, \cdots, N$.
$\bra{\tilde{\Psi}_0}$ and $\ket{\Psi_0}$ represent the left and right ground states of the system \cite{PhysRevResearch.2.033069,PhysRevB.83.245132}.

From Eq. \eqref{cab}, it's evident that the correlation matrix $C$ can be decomposed into:
\begin{equation}\label{correlation}
    C=\begin{pmatrix}
        Q_{AA}&Q_{AB}\\Q_{BA}&Q_{BB}
    \end{pmatrix}.
\end{equation}
Here,
\begin{align}
(Q_{AA})_{kj}=\bra{\tilde{\Psi}_0}\hat{a}^{\dagger}_k\hat{a}_j\ket{\Psi_0},&\quad(Q_{AB})_{kj}=\bra{\tilde{\Psi}_0}\hat{a}^{\dagger}_k\hat{b}_j\ket{\Psi_0},\nonumber\\(Q_{BA})_{kj}=\bra{\tilde{\Psi}_0}\hat{b}^{\dagger}_k\hat{a}_j\ket{\Psi_0},&\quad(Q_{BB})_{kj}=\bra{\tilde{\Psi}_0}\hat{b}^{\dagger}_k\hat{b}_j\ket{\Psi_0}.\nonumber
\end{align}

The remaining paragraph in Sec. \ref{nu} will focus on calculating these matrix elements using the Green's function we derived and comparing the results with direct numerical matrix decomposition.  

When $\ket{\Psi_0}=\otimes_{\varepsilon\in\mathds{E}}\ket{\psi_{\varepsilon}}$ and $\bra{\tilde{\Psi}_0}=\otimes_{\varepsilon\in\mathds{E}}\bra{\tilde{\psi}_{\varepsilon}}$, Eq. \eqref{53} can be alternatively expressed as:
\begin{equation}\label{co54}
    C=\sum_{\varepsilon\in\mathds{E}}\frac{\ket{\psi_{\varepsilon}}\bra{\tilde{\psi}_{\varepsilon}}}{\braket{\tilde{\psi}_{\varepsilon}|\psi_{\varepsilon}}}. 
\end{equation}
Comparing this equation with Eq. \eqref{G}, and using the residue theorem, we obtain:
\begin{equation}\label{c43}
    C=\frac{1}{2\pi i}\int_{\Gamma_{\mathds{E}}}G(\omega) d\omega.
\end{equation}
Here, $\Gamma_{\mathds{E}}$ is a closed curve containing all $\varepsilon$ in $\mathds{E}$. In conventional Hermitian systems, the ground state is commonly defined as the half-filled state, where particles occupy states from the lowest energy up to half of all available states. This implies completely filling one energy band in the case of two-band systems with chiral symmetry. However, in non-Hermitian systems, the notion of `half-filled' becomes ambiguous due to the presence of a complex energy spectrum \cite{Guo_2021,PhysRevB.105.205403}. In our study, we maintain the term `half-filled' to signify the occupation of one energy band, with further details provided later.

By substituting the expression of the Green's function of open boundary into Eq. \eqref{c43}, we find that the matrix elements in Eq. \eqref{correlation} can be given by:
\begin{small}
\begin{equation}\label{57}
\begin{split}
   (Q_{AA})_{kj}=\frac{1}{2}(q_{\delta}^{k-j}-\slashed{q}_{AA}^{k,j}),\quad& (Q_{AB})_{kj}=\frac{1}{2}(-q_{AB}^{k-j}+\slashed{q}_{AB}^{k,j});\\(Q_{BA})_{kj}=\frac{1}{2}(-q_{BA}^{k-j}+\slashed{q}_{BA}^{k,j}),\quad&(Q_{BB})_{kj}=\frac{1}{2}(q_{\delta}^{k-j}-\slashed{q}_{BB}^{k,j}).
   \end{split}
\end{equation}
\end{small}
for both the exact expression (Eq. \eqref{48}) and the formal expression (Eq. \eqref{eq48}). Refer to Appendix \ref{B} for detail deduction in Eq. \eqref{57}

The components present in Eq. \eqref{57} are different in the exact formula and the formal expression, we will employ numerical methods to validate them in both scenarios.
\subsection{The Exact Expression}
For the exact expression, the components in the Eq. \eqref{57} are given by: (Eq. \eqref{B19} and Eq. \eqref{B15})
\begin{align}\label{BB47}
 &q_{\delta}^{k-j}=\sum_z \frac{z^{k-j}}{2N_{\varepsilon}},\quad q_{AB}^{k-j}=\sum_z \frac{z^{k-j}}{2N_{\varepsilon}}q_{AB}(z),\nonumber\\&q_{BA}^{k-j}=\sum_z \frac{z^{k-j}}{2N_{\varepsilon}}q_{BA}(z),\quad\slashed{q}_{AA}^{k,j}=\sum_z\frac{z'^kz^{-j}}{2N_{\varepsilon}}q_{AB}(z')q_{BA}(z),\nonumber\\&\slashed{q}_{AB}^{k,j}=\sum_z\frac{z'^kz^{-j}}{2N_{\varepsilon}} q_{AB}(z'),\quad\slashed{q}_{BA}^{k,j}= \sum_z\frac{z'^kz^{-j}}{2N_{\varepsilon}} q_{BA}(z),\nonumber\\ &\slashed{q}_{BB}^{k,j}=\sum_z\frac{z'^kz^{-j}}{2N_{\varepsilon}}.
\end{align}
Here, $z'=r/z$,
\begin{equation}
\begin{split}
    q_{AB}(z)&=(t_1+\frac{\gamma}{2}+t_2z^{-1})/\varepsilon(z),\\ q_{BA}(z)&=(t_1-\frac{\gamma}{2}+t_2z)/\varepsilon(z).
\end{split}
\end{equation}
$N_{\varepsilon}=N-\frac{1}{4}\left[(q_1(z)q_2(z')+1)\braket{\Bar{z'}^{-1}|z}+(q_1(z')q_2(z)+1)\braket{\Bar{z}^{-1}|z'}\right]$, and $\varepsilon(z)$ is the eigenvalue of $h(z)$ in Eq. \eqref{13} that does not encircled by $\Gamma_{\mathds{E}}$.

\begin{figure}[h]
\centering
\includegraphics[width=1.0\linewidth]{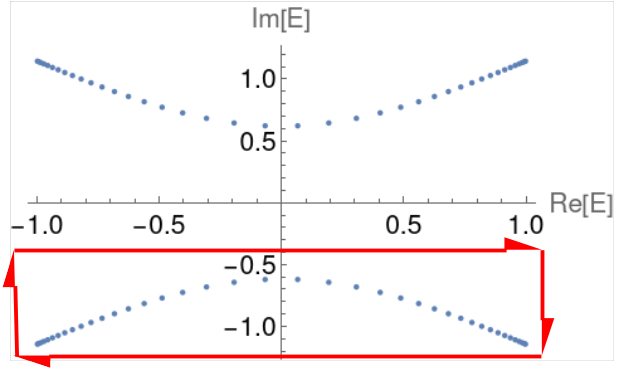}
\caption{The energy spectrum of open boundary non-Hermitian system, when $t_1=1/2,t_2=1,\gamma=5/2$,$N=40$. The colored closed curve is $\Gamma_{\mathds{E}}$, the arrow indicates the integral direction.}
\label{fig3}
\end{figure}

Let's take a specific example with $N=40$, $t_1=1/2$, $t_2=1$,$\gamma=5/2$ for numerical validation. Its energy spectrum is plotted in Fig. \ref{fig3}.

As observed in the plotted energy spectrum, the system exhibits symmetry around the real axis in the complex plane, and no edge states are present. This characteristic designates the current phase as the Entrapped Insulator, as illustrated in Fig. 15 of Ref. \cite{PhysRevB.105.205403}. In this phase, the energy spectrum is bifurcated into two bands—one with energy possessing a positive imaginary part and the other with energy having a negative imaginary part. As mentioned earlier, we define `half-filled’  as the occupation of one energy band. To be precise, we select the energy band with a negative imaginary part. In this case, $\Gamma_{\mathds{E}}$ is chosen to surround all energy points in the lower half-plane.

\begin{figure}[h]
\centering
\includegraphics[width=1.0\linewidth]{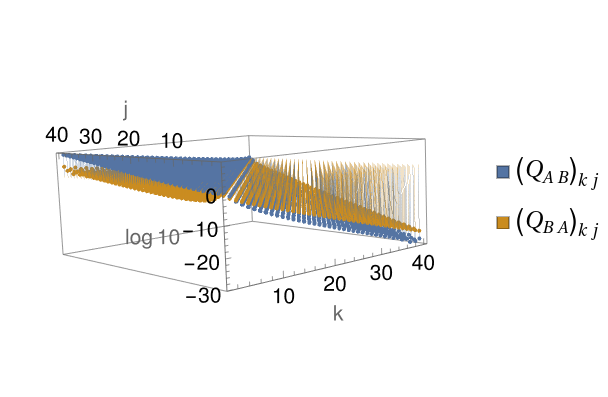}
\caption{\small The logarithm of absolute value of $(Q_{AB})_{kj}$ and $(Q_{BA})_{kj}$, when $N=40$, $t_1=1/2$, $t_2=1$,$\gamma=5/2$. The calculation results using the two methods are identical, the maximum relative error for (k,j) pairs $<10^{-32}$.}
\label{fig4}
\end{figure}

\begin{figure}[h]
\centering
\includegraphics[width=1.0\linewidth]{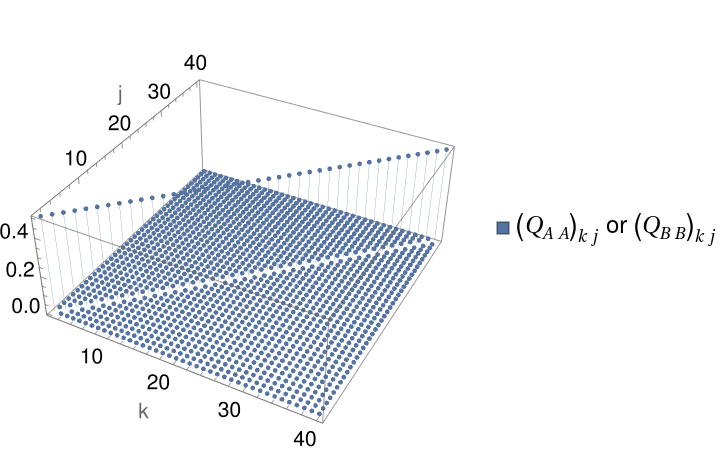}
\caption{\small The value of $(Q_{AA})_{kj}$ or $(Q_{BB})_{kj}$, When $N=40$, $t_1=1/2$, $t_2=1$,$\gamma=5/2$. The calculation results using the two methods are identical, the maximum relative error for (k,j) pairs $<10^{-32}$.}
\label{fig5}
\end{figure}

To verify Eq. \eqref{57}, we employ numerical methods. First, we perform an eigenvector decomposition of $H$, then use Eq. \eqref{co54} directly to obtain the correlation matrix. We compare this result with the one calculated using Eq. \eqref{57} and Eq. \eqref{BB47}. Remarkably, the results match precisely using both methods as depicted in Fig. \ref{fig4} and Fig. \ref{fig5}.

We use the relative error to measure the correspondence between our Green's function method and direct decomposition, defined as:
\begin{equation}
    <Q_{AA}>_{kj}=\frac{\lvert(Q_{AA}^g)_{kj}-(Q_{AA}^r)_{kj}\rvert}{\lvert(Q_{AA}^r)_{kj}\rvert}.
\end{equation}
Here, $<>$ denotes the relative error for the matrix we are concerned, superscript $^g$ denotes that this quantity is obtained by our Green's function method, and superscript $^r$ denotes that this quantity is obtained by direct decomposition.

Our calculation shows that for the exact expression we propose, ${\displaystyle\mathop{Max}_{1\leq k,j\leq N}}\{<Q_{AA}>_{kj},<Q_{AB}>_{kj},<Q_{BA}>_{kj},<Q_{BB}>_{kj}\}<10^{-32}
$, for the parameters we chose, suggesting they are identical.

Additionally, we observe that the results of $(Q_{AA})_{kj}$ and $(Q_{BB})_{kj}$ are precisely $\frac{1}{2}\delta_{kj}$ for Fig. \ref{fig5}. This behavior arises due to chirality. In Eq. \eqref{co54}, if the sum is taken for all $\varepsilon$ rather than just half, we obtain:
\begin{equation}
    \sum_{\varepsilon}\frac{\ket{\psi_{\varepsilon}}\bra{\tilde{\psi}_{\varepsilon}}}{\braket{\tilde{\psi}_{\varepsilon}|\psi_{\varepsilon}}}=\mathds{1}. 
\end{equation}
The existence of chirality implies:
\begin{small}
\begin{equation}
\begin{split}
    \sum_{\varepsilon}\frac{\ket{\psi_{\varepsilon}}\bra{\tilde{\psi}_{\varepsilon}}}{\braket{\tilde{\psi}_{\varepsilon}|\psi_{\varepsilon}}}=& \sum_{\varepsilon\in\mathds{E}}\frac{\ket{\psi_{\varepsilon}}\bra{\tilde{\psi}_{\varepsilon}}}{\braket{\tilde{\psi}_{\varepsilon}|\psi_{\varepsilon}}}+\sum_{\varepsilon\in\mathds{E}^c}\frac{\ket{\psi_{\varepsilon}}\bra{\tilde{\psi}_{\varepsilon}}}{\braket{\tilde{\psi}_{\varepsilon}|\psi_{\varepsilon}}}\\=&\sum_{\varepsilon\in\mathds{E}}\frac{\ket{\psi_{\varepsilon}}\bra{\tilde{\psi}_{\varepsilon}}}{\braket{\tilde{\psi}_{\varepsilon}|\psi_{\varepsilon}}}+\sum_{\varepsilon\in\mathds{E}}\sigma_z^{\otimes N}\frac{\ket{\psi_{\varepsilon}}\bra{\tilde{\psi}_{\varepsilon}}}{\braket{\tilde{\psi}_{\varepsilon}|\psi_{\varepsilon}}}\sigma_z^{\otimes N}.
    \end{split}
\end{equation}
\end{small}
Where $\sigma_z$ is the Pauli-z matrix. Evidently, we have:
\begin{small}
\begin{align}
   \bra{A,k} \sum_{\varepsilon\in\mathds{E}}\frac{\ket{\psi_{\varepsilon}}\bra{\tilde{\psi}_{\varepsilon}}}{\braket{\tilde{\psi}_{\varepsilon}|\psi_{\varepsilon}}}\ket{A,j}=\bra{A,k} \sum_{\varepsilon\in\mathds{E}}\sigma_z^{\otimes N}\frac{\ket{\psi_{\varepsilon}}\bra{\tilde{\psi}_{\varepsilon}}}{\braket{\tilde{\psi}_{\varepsilon}|\psi_{\varepsilon}}}\sigma_z^{\otimes N}\ket{A,j};\nonumber\\
   \bra{A,k} \sum_{\varepsilon\in\mathds{E}}\frac{\ket{\psi_{\varepsilon}}\bra{\tilde{\psi}_{\varepsilon}}}{\braket{\tilde{\psi}_{\varepsilon}|\psi_{\varepsilon}}}+\sum_{\varepsilon\in\mathds{E}}\sigma_z^{\otimes N}\frac{\ket{\psi_{\varepsilon}}\bra{\tilde{\psi}_{\varepsilon}}}{\braket{\tilde{\psi}_{\varepsilon}|\psi_{\varepsilon}}}\sigma_z^{\otimes N}\ket{A,j}=\delta_{kj}.
\end{align}
\end{small}

This implies $\bra{A,k}C\ket{A,j}=\frac{1}{2}\delta_{k,j}$, i.e., $(Q_{AA})_{kj}=\frac{1}{2}\delta_{kj}$. The same holds for $(Q_{BB})_{kj}$. Note that in the above deduction, we required $\mathds{E}$ to contain exactly half of all eigenstates. This is crucial as we are now considering situations with edge states.

\begin{figure}[h]
%\centering
\includegraphics[width=1.0\linewidth]{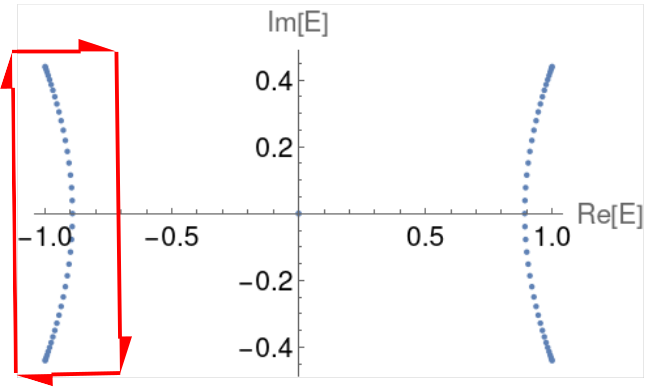}
\caption{The energy spectrum of open boundary non-Hermitian system, when $N=40, t_1=1/2,t_2=1,\gamma=4/3$. The colored closed curve is $\Gamma_{\mathds{E}}$, the arrow indicates the integral direction.}
\label{fig6}
\end{figure}

\begin{figure}[h]
\centering
\includegraphics[width=1.0\linewidth]{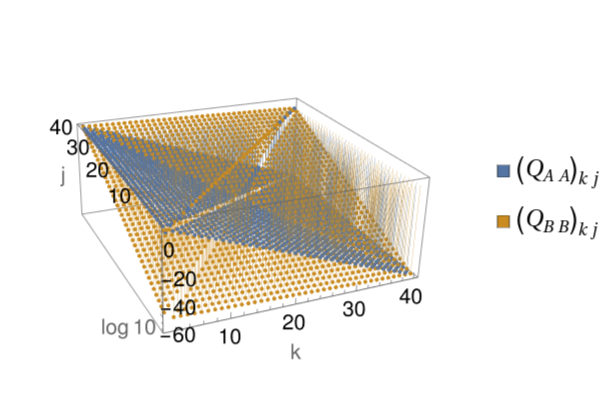}
\caption{\small The logarithm of absolute value of $(Q_{AA})_{kj}$ and $(Q_{BB})_{kj}$, when $N=40, t_1=1/2,t_2=1,\gamma=4/3$. The calculation results using the two methods are identical, the maximum relative error for (k,j) pairs $<1.3 \times 10^{-7}$.}
\label{fig7}
\end{figure}

\begin{figure}[h]
\centering
\includegraphics[width=1.0\linewidth]{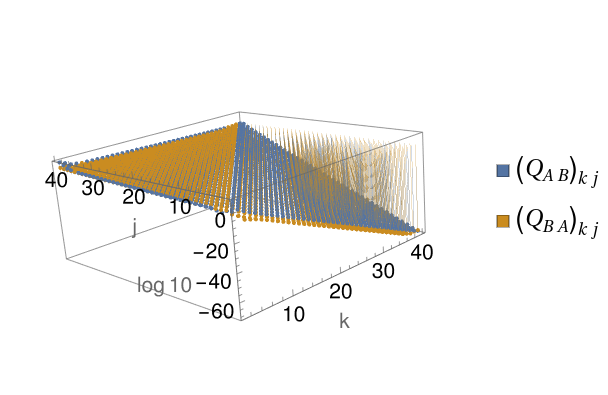}
\caption{\small The logarithm of absolute value of $(Q_{AB})_{kj}$ and $(Q_{BA})_{kj}$, when $N=40, t_1=1/2,t_2=1,\gamma=4/3$. The calculation results using the two methods are identical, the maximum relative error for (k,j) pairs $<1.3 \times 10^{-7}$.}
\label{fig8}
\end{figure}

Consider a system with $N=40, t_1=1/2,t_2=1,\gamma=4/3$, its energy spectrum is plotted in Fig. \ref{fig6}.
In this scenario, the energy spectrum exhibits symmetry around the imaginary axis and features edge states, designating this phase as the Topological Insulator phase, as illustrated in Fig. 15 of ref \cite{PhysRevB.105.205403}. Notably, in this case, the energy spectrum is divided into two bands: one with energy exhibiting a positive real part, and the other with energy having a negative real part. In our definition of the ground state for this case, we fill up the energy band characterized by a negative real part. In other words, $\Gamma_{\mathds{E}}$ is selected to envelop all energy points situated in the left half-plane. The numerical results of the correlation matrix are depicted in Fig. \ref{fig7} and \ref{fig8}.

The results obtained using equations Eq. \eqref{57}, and Eq. \eqref{BB47} are consistent with the direct decomposition. The calculation shows that ${\displaystyle\mathop{Max}_{1\leq k,j\leq N}}\{<Q_{AA}>_{kj},<Q_{AB}>_{kj},<Q_{BA}>_{kj},<Q_{BB}>_{kj}\}<1.3\times10^{-7}$ in this parameter set. It's important to note that $(Q_{AA})_{kj}$ and $(Q_{BB})_{kj}$ in Fig .\ref{fig7} are not precisely $\frac{1}{2}\delta_{kj}$ due to the presence of edge states. The set $\mathds{E}$ that we are concerned with is not exactly half of all states. However, including one of the edge states in Eq. \eqref{co54} allows us to recover $\frac{1}{2}\delta_{kj}$ for $(Q_{AA})_{kj}$ and $(Q_{BB})_{kj}$.
\subsection{The Formal Expression}
In the final analysis, we aim to validate our formal theory with numerical results. For the formal expression, the components in the Eq. \eqref{57} are given by: (Eq. \eqref{B11} and Eq. \eqref{Blast})
\begin{widetext}
\begin{equation}\label{BB53}
 q_{\delta}^{k-j}=\delta_{k,j},\quad q_{AB}^{k-j}=\int_{|z|\equiv\sqrt{|r|}}\frac{dz}{2\pi i z} q_{AB}(z)z^{k-j},\quad q_{BA}^{k-j}=\int_{|z|\equiv\sqrt{|r|}}\frac{dz}{2\pi i z}q_{BA}(z)z^{k-j},
\end{equation}
 \begin{small}
 \begin{align}
 &\slashed{q}_{AA}^{k,j}=\int_{|z|\equiv \sqrt{|r|}}\frac{dz}{2\pi i z}\left(q_{AB}(z')q_{BA}(z)z'^kz^{-j}+z'^{(k-N-1)}z^{(N+1-j)}\right),\quad\slashed{q}_{AB}^{k,j}=\int_{|z|\equiv \sqrt{|r|}}\frac{dz}{2\pi i z}\left(q_{AB}(z')z'^kz^{-j}+q_{AB}(z)z'^{(k-N-1)}z^{(N+1-j)}\right),\nonumber\\&\slashed{q}_{BA}^{k,j}=\int_{|z|\equiv \sqrt{|r|}}\frac{dz}{2\pi i z}\left(q_{BA}(z)z'^kz^{-j}+q_{BA}(z')z'^{(k-N-1)}z^{(N+1-j)}\right),\quad\slashed{q}_{BB}^{k,j}=\int_{|z|\equiv \sqrt{|r|}}\frac{dz}{2\pi i z}\left(z'^kz^{-j}+q_{AB}(z)q_{BA}(z')z'^{(k-N-1)}z^{(N+1-j)}\right).\nonumber
\end{align}
\end{small}
\end{widetext}
We still chose the system with $N=40, t_1=1/2,t_2=1,\gamma=4/3$, the relative error between formal theory (Eq. \eqref{57} and Eq. \eqref{BB53}) and the numerical results is shown in Fig. \ref{fig9} and Fig. \ref{fig10}:
\begin{figure}[h]
\centering
\includegraphics[width=1.0\linewidth]{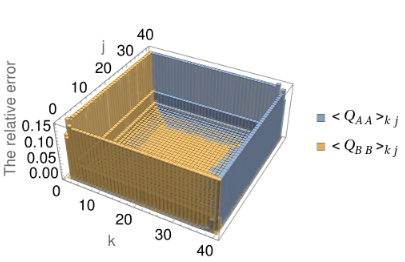}
\caption{\small The relative error of $(Q_{AA})_{kj}$ and $(Q_{BB})_{kj}$ between formal theory and numerical results,  when $N=40, t_1=1/2,t_2=1,\gamma=4/3$. Choose 360 points uniformly in circle $|z|=\sqrt{|r|}$ when doing numerical integration of formal theory.}
\label{fig9}
\end{figure}
\begin{figure}[h]
\centering
\includegraphics[width=1.0\linewidth]{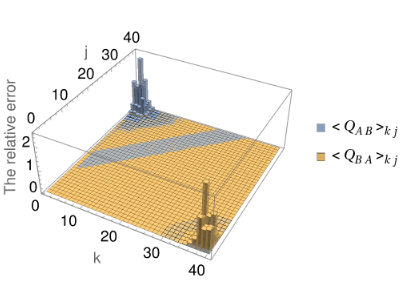}
\caption{\small The relative error of $(Q_{AB})_{kj}$ and $(Q_{BA})_{kj}$ between formal theory and numerical results, when $N=40, t_1=1/2,t_2=1,\gamma=4/3$. Choose 360 points uniformly in circle $|z|=\sqrt{|r|}$ when doing numerical integration of formal theory.}
\label{fig10}
\end{figure}

The agreement between our formal theory and numerical results, as shown in Fig. \ref{fig9} and Fig. \ref{fig10}, is remarkable. Our formal expression can describe the system well in the regime away from the boundary. Some points near the edge are mismatched because the integrated expression does not converge in this regime as discussed in Sec. \ref{sec:4}. However, for $Q_{AA}$ and $Q_{BB}$, even in regions close to the edge, where the relative error is under $20\%$, our formal theory provides a consistent description of $Q_{AA}$ and $Q_{BB}$. Disregarding this relatively small error, we can express $Q_{AA}$ and $Q_{BB}$ as follows:
\begin{equation}\label{eq62}
\begin{split}
    (Q_{AA})_{kj}=&\frac{1}{2}\left(q_{\delta}^{k-j}+\slashed{q}_{AA}^{k,j}\right)=\frac{1}{2}\left(\delta_{kj}+\slashed{q}_{AA}^{k,j}\right),\\
    (Q_{BB})_{kj}=&\frac{1}{2}\left(q_{\delta}^{k-j}+\slashed{q}_{BB}^{k,j}\right)=\frac{1}{2}\left(\delta_{kj}+\slashed{q}_{BB}^{k,j}\right).
\end{split}
\end{equation}
Here, we use the fact that $q_{\delta}^{k-j}=\delta_{kj}$ in Eq. \eqref{BB53}. As mentioned earlier, including one of the terms of the edge state in the correlation matrix will recover $\frac{1}{2}\delta_{kj}$. This is expressed as:
\begin{equation}
 \begin{split}
     (Q_{AA})_{kj}+\bra{A,k}\left(\ket{0_1}\bra{\tilde{0}_1}\right)\ket{A,j}=\frac{1}{2}\delta_{kj},\\   (Q_{BB})_{kj}+\bra{B,k}\left(\ket{0_1}\bra{\tilde{0}_1}\right)\ket{B,j}=\frac{1}{2}\delta_{kj}.
 \end{split}   
\end{equation}
    Because of Eq. \eqref{eq62}, we can have:
\begin{equation}\label{e56}
\begin{split}
    \bra{A,k}\left(\ket{0_1}\bra{\tilde{0}_1}\right)\ket{A,j}=-\frac{1}{2}\slashed{q}_{AA}^{k,j},\\
    \bra{B,k}\left(\ket{0_1}\bra{\tilde{0}_1}\right)\ket{B,j}=-\frac{1}{2}\slashed{q}_{BB}^{k,j}.
\end{split}
\end{equation} 

Where $\slashed{q}_{AA}^{k,j}$ and $\slashed{q}_{BB}^{k,j}$ submission to Eq. \eqref{BB53}. This result is particularly interesting as it provides insights into the behavior of the edge state in the context of the correlation matrix. 

We further note that because of chirality, 
\begin{align}
    &\ket{0_2}\bra{\tilde{0}_2}=\sigma_z^{\otimes N}\ket{0_1}\bra{\tilde{0}_1}\sigma_z^{\otimes N},\\
    \Rightarrow&
\left\{\begin{aligned}
    &\bra{A,k}\left(\ket{0_1}\bra{\tilde{0}_1}\right)\ket{A,j}=\bra{A,k}\left(\ket{0_2}\bra{\tilde{0}_2}\right)\ket{A,j};\\
   & \bra{B,k}\left(\ket{0_1}\bra{\tilde{0}_1}\right)\ket{B,j}=\bra{B,k}\left(\ket{0_2}\bra{\tilde{0}_2}\right)\ket{B,j}.   
\end{aligned}\right.\nonumber
\end{align}
Using Eq. \eqref{e39}, we can rewrite Eq. \eqref{e56} as:
\begin{equation}\label{ee52}
\begin{split}
    \bra{A,k}G_{edge}(\omega)\ket{A,j}=-\frac{1}{\omega}\slashed{q}_{AA}^{k,j},\\\bra{B,k}G_{edge}(\omega)\ket{B,j}=-\frac{1}{\omega}\slashed{q}_{BB}^{k,j}.
\end{split}
\end{equation}
\section{Summary and Discussion}
In this paper, starting from the generalized Bloch theorem, we have derived the generalized Brillouin Zone theory. In particular, we examined a classical non-Hermitian model under periodic and open boundary conditions. We have successfully derived exact summation expressions for both scenarios (Eq. \eqref{eq37} for periodic boundary conditions and Eq. \eqref{eq47} for open boundary conditions). Additionally, we have extended our findings to encompass an infinite-size system under open boundary conditions, presenting a formal expression in integral form (Eq. \eqref{eq48}). Notably, in comparison to existing literature \cite{hu2023greens, PhysRevB.103.L241408,PhysRevB.107}, the term $G_{bound}$ in Eq. \eqref{eq48} can be viewed as a correction to their formulas. 
We have also introduced a method for deriving the correlation matrix using the Green function. We validated our theoretical framework by comparing numerical results with outcomes derived from our approach. The numerical analysis not only confirms the accuracy of our exact formula but also demonstrates the high fidelity of our formal expression.

The theoretical framework proposed in this paper boasts versatile applications. Beyond the conventional use of Green's function for system response and dynamic description, its relationship with the correlation function allows for seamless application in deriving and theoretically analyzing entanglement spectra and the system's topological properties. An illustrative example showcasing its utility in edge states has been provided in Eq. \eqref{ee52}, hinting at its potential to offer new insights into bulk-boundary correspondence in non-Hermitian systems.  Although our investigation focused on a model with chiral symmetry, we posit that the framework we proposed applies to a broader family of non-Hermitian systems, encompassing various boundary conditions.

One limitation of this paper lies in our numerical tests, which were confined to a lattice of size 40. Future work could delve into the scaling behavior of our expressions, spanning from finite to infinite lattice sizes, providing a more comprehensive understanding of their applicability.
\begin{acknowledgments}
%The authors would like to thank ChatGPT and Grammarly for proofreading and polishing our draft.
This work was supported by the National Natural Science Foundation of China under Grant No. 12175015.
\end{acknowledgments}
\appendix
\section{Some Deduction Details of the Green's Function in Sec. \ref{sec:4}\label{A}}
For a system with Hamiltonian $H$, its Green's function can be expressed in formalization as:
\begin{equation}
    G(\omega)=\frac{1}{\omega-H}.
\end{equation}
For the Hermitian system, the above equation can be further written as:
\begin{equation}\label{11}
    G(\omega)=\sum_n\frac{\ket{\phi_n}\bra{\phi_n}}{\omega-H}=\sum_n\frac{\ket{\phi_n}\bra{\phi_n}}{\omega-E_n}.
\end{equation}
Where $\{\ket{\phi_n}\}$ is the right eigenstates of $H$ satisfy $H\ket{\phi_n}=E_n\ket{\phi_n}$ and $\sum_n\ket{\phi_n}\bra{\phi_n}=\mathds{1}$, $\braket{\phi_n|\phi_m}=\delta_{nm}$. However, for a non-Hermitian system, its right eigenstates do not have orthogonality, usually, we choose both left and right eigenstates to form an orthogonal complete set \cite{Brody_2014}, i.e.
\begin{align}
    H\ket{\phi_n}=E_n\ket{\phi_n},\quad \bra{\chi_n}H=E_n\bra{\chi_n};\\
    \braket{\chi_n|\phi_m}=\braket{\chi_n|\phi_n}\delta_{nm},\quad\sum_n\frac{\ket{\phi_n}\bra{\chi_n}}{\braket{\chi_n|\phi_n}}=\mathds{1}.
\end{align}
Hence, the Green function in the non-Hermitian system can be further written as:
\begin{equation}\label{above}
    G(\omega)=\sum_n\frac{\ket{\phi_n}\bra{\chi_n}}{\braket{\chi_n|\phi_n}}\frac{1}{\omega-H}=\sum_n\frac{\ket{\phi_n}\bra{\chi_n}}{\braket{\chi_n|\phi_n}}\frac{1}{\omega-E_n}.
\end{equation}
For systems we concerned, just substitute Eq. \eqref{con} into the Eq. \eqref{above}, we can get:
\begin{equation}\label{AG}
    G(\omega)=\sum_{\varepsilon}\frac{\ket{\psi_{\varepsilon}}\bra{\Tilde{\psi_{\varepsilon}}}}{\braket{\Tilde{\psi_{\varepsilon}}|\psi_{\varepsilon}}}\frac{1}{\omega-\varepsilon},
\end{equation}
    where $\bra{\Tilde{\psi_{\varepsilon}}}$ is the correspond left eigenvector of $\ket{\psi_{\varepsilon}}$, satisfies:
\begin{equation}    \bra{\Tilde{\psi_{\varepsilon}}}H=\varepsilon\bra{\Tilde{\psi_{\varepsilon}}}\Rightarrow H^{\dagger}\ket{\Tilde{\psi_{\varepsilon}}}=\bar{\varepsilon}\ket{\Tilde{\psi_{\varepsilon}}}.
\end{equation}
Since we have $\bra{i}H^{\dagger}\ket{j}=h_{i-j}^{\dagger}$, its correspond bulk Hamiltonian is $\Tilde{h}(z)=\sum_{l=-r}^{r}h^{\dagger}_{-l}z^l=(h(\bar{z}^{-1}))^{\dagger}$, if $\Tilde{h}(z)$ has a right eigenvector $\widetilde{\ket{u(z)}}$ satisfies:
\begin{equation}
    \Tilde{h}(z)\widetilde{\ket{u(z)}}=\bar{\varepsilon}\widetilde{\ket{u(z)}}\iff (h(\bar{z}^{-1}))^{\dagger}\widetilde{\ket{u(z)}}=\bar{\varepsilon}\widetilde{\ket{u(z)}},
\end{equation}
then $\widetilde{\bra{u(z)}}$ is the left eigenvector of $h(\bar{z}^{-1})$ for eigenvalue $\varepsilon$. If we denote the left eigenvector of $h(z)$ for $\varepsilon$ as $\bra{\Tilde{u}(z)}$, then $\widetilde{\bra{u(z)}}=\bra{\Tilde{u}(\bar{z}^{-1})}$, then the left eigenvector of $H$ for $\varepsilon$ is $\widetilde{\bra{u(z)}}\bra{z}=\bra{\Tilde{u}(\bar{z}^{-1})}\bra{z}$ or $\bra{\Tilde{u}(z)}\bra{\bar{z}^{-1}}$. We note here that one can always rescale $\bra{\Tilde{u}(z)}$ to satisfy $\braket{\Tilde{u}(z)|u(z)}=1$ just for convince, and we will make it as default in the later calculation.

We can further check this assert by using the translational operator $T$, since $\ket{z}=\sum_{j\in\mathds{Z}}z^j\ket{j}$, we have:
\begin{equation}\label{18}
    \ket{\bar{z}^{-1}}=\sum_{j\in\mathds{Z}}\bar{z}^{-j}\ket{j},\quad \bra{\bar{z}^{-1}}=\sum_{j\in\mathds{Z}}z^{-j}\bra{j},\\
 \end{equation}  
\begin{equation}\label{19}
    \bra{\bar{z}^{-1}} T=\sum_{j\in\mathds{Z}}z^{-j}\bra{j} \sum_{j\in\mathds{Z}}\ket{j}\bra{j+1}= \bra{\bar{z}^{-1}}z.
 \end{equation}
in correspond to Eq. \eqref{4}. Therefore, we can deduce that $\bra{\Tilde{u}(z)}\bra{\bar{z}^{-1}}$ is the correspond left eigenvector of $\ket{u(z)}\ket{z}$.

Let's return to Eq. \eqref{AG}, our goal is to derive the Green function. The periodic condition is relatively easier since its wavefunction only consists of one component, substituting Eq. \eqref{eq18} and its corresponding left eigenvector into Eq. \eqref{AG}, 
we can get:
\begin{equation}
    G(\omega)=\sum_{\varepsilon}\frac{\ket{u(z)}\bra{\tilde{u}(z)}}{N(\omega-\varepsilon)}\ket{z}\bra{\bar{z}^{-1}}.
\end{equation} 
This is the Green's function of open boundary, changing the subscript in the sum from $\varepsilon$ into $z$, yields Eq. \eqref{eq37}. 

Move on to the open boundary condition next. Denote the corresponding left eigenvector of Eq. \eqref{eq25} as:
\begin{equation}\label{A40}\bra{\tilde{\psi}_{\varepsilon}}=\tilde{c}_{z}\bra{\tilde{u}(z)}\bra{\bar{z}^{-1}}+\tilde{c}_{z'}\bra{\tilde{u}(z')}\bra{\bar{z'}^{-1}}.
\end{equation}
There is no reason to assume the coefficient $\tilde{c}_{z}$ and $\tilde{c}_{z'}$ are the conjugate of $c_z$ and $c_{z'}$ like hermitian case. Instead, we shall use the boundary matrix again to decide them.

Since the left eigenvector of $H$ is the right eigenvector of $H^{\dagger}$, so its boundary matrix yields:
\begin{equation}
\begin{split}
\tilde{B}(\varepsilon)
    =&\begin{pmatrix}
     P_{\partial}(H^{\dagger}-\bar{\varepsilon}\mathds{1})\ket{\tilde{u}(z)}\ket{\bar{z}^{-1}} \\  P_{\partial}(H^{\dagger}-\bar{\varepsilon}\mathds{1})\ket{\tilde{u}(z')}\ket{\bar{z'}^{-1}}
    \end{pmatrix}^T\\=&\begin{pmatrix}
        -h_1^{\dagger}\ket{\tilde{u}(z)}&-h_1^{\dagger}\ket{\tilde{u}(z')}\\-\bar{z}^{-(N+1)}h_{-1}^{\dagger}\ket{\tilde{u}(z)}&-\bar{z'}^{-(N+1)}h_{-1}^{\dagger}\ket{\tilde{u}(z')}
    \end{pmatrix}.
\end{split}
\end{equation}
Further denote:
\begin{equation}
    \bra{\tilde{u}(z)}=\begin{pmatrix}
   \tilde{\phi}_A\\\tilde{\phi}_B    
    \end{pmatrix}^T,\quad\bra{\tilde{u}(z')}=\begin{pmatrix}
   \tilde{\phi}_A'\\\tilde{\phi}_B'
    \end{pmatrix}^T.
\end{equation}

We can get:
\begin{equation}
    \frac{\tilde{c}_z}{\tilde{c}_{z'}}=-\frac{\tilde{\phi}_B'}{\tilde{\phi}_B}\quad\text{or}\quad \frac{\tilde{c}_z}{\tilde{c}_{z'}}=-\frac{z'^{-(N+1)}\tilde{\phi}_A'}{z^{-(N+1)}\tilde{\phi}_A},
\end{equation}
similar to Eq. \eqref{ab26}. 

Denote $\braket{\Tilde{\psi_{\varepsilon}}|\psi_{\varepsilon}}=N_{\varepsilon}$, $N_{\varepsilon}$ will go to N as N increase to $\infty$.
This is because $z$ and $z'$ have the same module in our model. Assume $z=\sqrt{|r|}e^{i\theta}$, $z'=\sqrt{|r|}e^{i\theta'}$, we would have:
\begin{small}
\begin{equation}
    \frac{1}{N}\braket{\bar{z'}^{-1}|z}=\frac{1}{N}\sum_j^Ne^{i(\theta-\theta')j}=\frac{1}{N}\frac{1-e^{i(\theta-\theta')(N+1)}}{1-e^{i(\theta-\theta')}},
\end{equation}
\end{small}
\begin{small}
\begin{equation}
    \frac{1}{N}\left|\frac{1-e^{i(\theta-\theta')(N+1)}}{1-e^{i(\theta-\theta')}}\right|<\frac{2}{N}\frac{1}{|1-e^{i(\theta-\theta')}|}\xrightarrow[\theta\neq\theta']{N\to\infty}0,
\end{equation}
\end{small}
when $N\to\infty$. Resulting:
\begin{equation}
\braket{\bar{z}^{-1}|z'},\braket{\bar{z'}^{-1}|z}\to N\delta_{zz'}.
\end{equation}
Therefore, $N_{\varepsilon}\to N$ in limit case.

Substituting Eq. \eqref{A40} and Eq. \eqref{eq25} into Eq. \eqref{AG}, we derive the Green's function:
\begin{small}
\begin{equation}
\begin{split}
    &G(\omega)=\sum_{\varepsilon}\frac{1}{N_{\varepsilon}(\omega-\varepsilon)}\Big(c_z\tilde{c}_{z}\ket{u(z)}\bra{\Tilde{u}(z)}\ket{z}\bra{\bar{z}^{-1}}+\\&c_{z'}\tilde{c}_{z'}\ket{u(z')}\bra{\Tilde{u}(z')}\ket{z'}\bra{\bar{z'}^{-1}}+c_z\tilde{c}_{z'}\ket{u(z)}\bra{\tilde{u}(z')}\ket{z}\bra{\bar{z'}^{-1}}\\&+\tilde{c}_{z}c_{z'}\ket{u(z')}\bra{\tilde{u}(z)}\ket{z'}\bra{\bar{z}^{-1}}\Big),
\end{split}
\end{equation}
\end{small}

We now aim to change the sum over $\varepsilon$ to the sum over $z$ again. However, as a $z$ goes over the GBZ, its corresponding $z'$ will also go around the GBZ. Therefore, $\sum_{\varepsilon}\to\frac{1}{2}\sum_l\sum_{z}$, and we obtain:
\begin{small}
\begin{equation}\label{A47}
\begin{split}
    &G(\omega)=\frac{1}{2}\sum_l\sum_{z}\frac{1}{N_{\varepsilon}(\omega-\varepsilon_l(z))}\Big(c_z\tilde{c}_{z}\ket{u_l(z)}\bra{\Tilde{u}_l(z)}\ket{z}\bra{\bar{z}^{-1}}+\\&c_{z}\tilde{c}_{z}\ket{u_l(z)}\bra{\Tilde{u}_l(z)}\ket{z}\bra{\bar{z}^{-1}}+c_{z'}\tilde{c}_{z}\ket{u_l(z')}\bra{\tilde{u}_l(z)}\ket{z'}\bra{\bar{z}^{-1}}\\&+\tilde{c}_{z}c_{z'}\ket{u_l(z')}\bra{\tilde{u}_l(z)}\ket{z'}\bra{\bar{z}^{-1}}\Big).
\end{split}
\end{equation}
\end{small}

We have made the second term in the sum more symmetrical by changing $z'$ to $z$ and switching the order of $z$ and $z'$ in the third term. If we chose $c_z\tilde{c}_z+c_{z'}\tilde{c}_{z'}=1$ for convenience, Eq. \eqref{A47} can be simplified to:
\begin{equation}\label{A21}
\begin{split}
    G(\omega)=&\frac{1}{2}\sum_l\sum_{z}\frac{1}{N_{\varepsilon}(\omega-\varepsilon_l(z))}\Big(\ket{u_l(z)}\bra{\Tilde{u}_l(z)}\ket{z}\bra{\bar{z}^{-1}}\\+&2c_{z'}\tilde{c}_{z}\ket{u_l(z')}\bra{\tilde{u}_l(z)}\ket{z'}\bra{\bar{z}^{-1}}\Big).
\end{split}
\end{equation}

Incorporating Eq. \eqref{A21} into a more generalized system with open boundary conditions is possible. In \cite{2021,PhysRevLett.123.066404}, the authors emphasize that in a non-Hermitian system under open boundaries, wavefunctions consistently appear in the form $(z, z')$, where $|z| = |z'|$. This phenomenon arises due to the requirement of the wavefunction to be a stationary wave and vanish at the open boundary, necessitating two parts with equal intensity. 

This indicates that the wavefunction would closely resemble Eq. \eqref{eq25}, except for the specific GBZ. Since $|z| = |z'|$, the property $N_{\varepsilon} \to N$ is preserved, suggesting that its formal expression for infinite size remains the same as Eq. \eqref{eq48}, at least when all sub-GBZs are circular. However, if the sub-GBZs are not circular, Eq. \eqref{p49} may not be satisfied, as is possible in certain multiband systems \cite{PhysRevLett.125.226402}.

In such cases, the difference of $z$ can be expressed as follows:
\begin{equation}
\Delta z \approx \frac{\pi iz}{N} + \Delta |z|.
\end{equation}

Catastrophically, the integral can only be expressed as $\displaystyle\int_0^{2\pi}\frac{d\theta}{2\pi}$, where $\theta$ is the main argument of $z$, given by $z=re^{i\theta}$. Since $\displaystyle   \oint_z\frac{dz}{2\pi i z}=\oint_z\frac{dr}{2\pi i r}+\int_0^{2\pi}\frac{d\theta}{2\pi}$, a potential adjustment to the formal expression is to replace
$\displaystyle
    \oint_{z\in GBZ_l} \frac{dz}{2\pi i z}\quad\text{with}\quad \oint_{z\in GBZ_l} \left(\frac{dz}{2\pi i z }-\frac{d|z|}{2\pi i |z| }\right)$, or simply written as $\displaystyle \oint_{z\in GBZ_l} \frac{d\theta}{2\pi}$.
\section{Deduction Details in Eq. \eqref{57} \label{B}}
In this appendix, we are going to deduct the matrix element in Eq. \eqref{correlation} using Eq. \eqref{c43}. In Eq. \eqref{ee40}, we see that $G(\omega)$ has three parts, we should consider the integral of $G_{bulk}$ in the exact expression (Eq. \eqref{48}) first.

Follow the denotation in Eq. \eqref{u21} and Eq. \eqref{u42}, and replace the subscribe of $l$ into $\pm$. Because of chiral symmetry, we have:
\begin{equation}
    \ket{u_{\pm}(z)}=\begin{pmatrix}
        \phi_A\\\pm\phi_B
    \end{pmatrix},\quad \bra{\Tilde{u}_{\pm}(z)}=\begin{pmatrix}
        \tilde{\phi}_A\\\pm\tilde{\phi}
        _B
    \end{pmatrix}^T.
\end{equation}
Here,
\begin{equation}\label{B1}
\begin{split}  
    h(z)\ket{u_{\pm}(z)}=\pm\varepsilon(z)\ket{u_{\pm}(z)},\\\bra{\Tilde{u}_{\pm}(z)}h(z)=\pm\varepsilon(z)\bra{\Tilde{u}_{\pm}(z)}.
\end{split}
\end{equation}
Therefore,
\begin{equation}\label{B2}
\left\{
\begin{aligned}
    &\braket{\Tilde{u}_{\pm}(z)|u_{\pm}(z)}=1,\\&\braket{\Tilde{u}_{\pm}(z)|u_{\mp}(z)}=0,
    \end{aligned}
    \right.
    \Rightarrow\phi_A\Tilde{\phi}_A=\phi_B\Tilde{\phi}_B=\frac{1}{2}.
\end{equation}
This equation is very important and will be used thoroughly. Moreover:
\begin{equation}
\begin{split}
    \ket{u_{\pm}(z)}\bra{\Tilde{u}_{\pm}(z)}&=\begin{pmatrix}
        \phi_A\tilde{\phi}_A&\pm\phi_A\tilde{\phi}_B\\\pm\phi_B\tilde{\phi}_A&\phi_B\tilde{\phi}_B
    \end{pmatrix}\\&=\frac{1}{2}\begin{pmatrix}
        1&\pm q_{AB}(z)\\\pm q_{BA}(z)&1
    \end{pmatrix},
 \end{split}   
\end{equation}
where we denote $q_{AB}(z)=2\phi_A\tilde{\phi}_B$, $q_{BA}(z)=2\phi_B\tilde{\phi}_A$. 

Because of Eq. \eqref{B1}, we have:
\begin{equation}\label{B5}
\begin{split}
&\left\{\begin{aligned}
    &(t_1+\frac{\gamma}{2}+t_2z^{-1})\phi_B=\varepsilon\phi_A;\\
    &(t_1-\frac{\gamma}{2}+t_2z)\phi_A=\varepsilon\phi_B.   
\end{aligned}\right.\\\Rightarrow&
\left\{\begin{aligned}
    &(t_1+\frac{\gamma}{2}+t_2z^{-1})/2=\varepsilon\phi_A\tilde{\phi}_B;\\
   & (t_1-\frac{\gamma}{2}+t_2z)/2=\varepsilon\phi_B\tilde{\phi}_A.   
\end{aligned}\right.
\end{split}
\end{equation}
i.e.
\begin{equation}
\begin{split}
    q_{AB}(z)&=(t_1+\frac{\gamma}{2}+t_2z^{-1})/\varepsilon,\\ q_{BA}(z)&=(t_1-\frac{\gamma}{2}+t_2z)/\varepsilon.
\end{split}    
\end{equation}

We use Eq. \eqref{B2} to deduct Eq. \eqref{B5}. Return to Eq. \eqref{48}, we have:
\begin{equation}\label{B7}
\begin{split}
    &\sum_l\frac{\ket{u_{l}(z)}\bra{\Tilde{u}_{l 
    }(z)}}{\omega-\varepsilon_l(z)}\\=&\frac{1}{2}\frac{1}{\omega^2-\varepsilon^2(z)}\begin{pmatrix}
       \omega&\varepsilon(z)q_{AB}(z)\\\varepsilon(z)q_{BA}(z)&\omega 
    \end{pmatrix}.
\end{split}
\end{equation}
The subsequent step involves integrating around $\Gamma_{\mathds{E}}$. Due to the chirality, the two bands of the energy spectrum display origin symmetry in the complex plane. Given our selection of the ground state, filled from one energy band, $\Gamma_{\mathds{E}}$ should encircle one energy band according to our choice. Assuming we denote all the energy points in the chosen band as $-\varepsilon$, then $\Gamma_{\mathds{E}}$ should surround all $-\varepsilon$s.

By substituting Eq. \eqref{B7} into Eq. \eqref{48} and integrating around $\Gamma_{\mathds{E}}$, the result becomes:
\begin{equation}\label{B8}
\begin{split}
    (C_{bulk})_{kj}=&\sum_z\frac{z^{k-j}}{2N_{\varepsilon}}\int_{\Gamma_{\mathds{E}}}\frac{d\omega}{2\pi i}\sum_l\frac{\ket{u_{l}(z)}\bra{\Tilde{u}_{l 
    }(z)}}{\omega-\varepsilon_l(z)}\\
    =&\sum_z\frac{z^{k-j}}{2N_{\varepsilon}}\frac{1}{2} 
    \begin{pmatrix}
        1&-q_{AB}(z)\\-q_{BA}(z)&1
    \end{pmatrix}. 
\end{split}
\end{equation}
Here, we denote $C_{bulk}$ as the correlation matrix corresponding to $G_{bulk}$. $C_{bulk}$ can also be written as:
\begin{equation}
   C_{bulk}= \frac{1}{2}\begin{pmatrix}
        q_{\delta}^{k-j}&-q_{AB}^{k-j}\\-q_{BA}^{k-j}&q_{\delta}^{k-j}
    \end{pmatrix},
\end{equation}
if we denote:
\begin{equation}\label{B19}
\begin{split}
 &q_{\delta}^{k-j}=\sum_z \frac{z^{k-j}}{2N_{\varepsilon}},\quad q_{AB}^{k-j}=\sum_z \frac{z^{k-j}}{2N_{\varepsilon}}q_{AB}(z),\\ &q_{BA}^{k-j}=\sum_z \frac{z^{k-j}}{2N_{\varepsilon}}q_{BA}(z).
 \end{split}
\end{equation}

We are going to consider the integral over $G_{bound}$ next. First is to calculate the coefficient $2c_{z'}\tilde{c}_{z}$. Because of Eq. \eqref{ab26} and Eq. \eqref{eq43}, we can assume:
\begin{equation}
    c_z=k\phi_B',\quad c_{z'}=-k\phi_B;\quad
    \tilde{c}_z=l\tilde{\phi}_B',\quad\tilde{c}_{z'}=-l\tilde{\phi}_B. 
\end{equation}
Since we have  $c_z\tilde{c}_z+c_{z'}\tilde{c}_{z'}=1$ and Eq. \eqref{B2}, we have $kl=1$ and:
\begin{equation}\label{cz1}
    2c_{z'}\tilde{c}_{z}=-2\phi_B\tilde{\phi}_B'.
\end{equation}

Or we can assume:
\begin{equation}
\begin{split}
    c_z=kz'^{N+1}\phi_A',&\quad c_{z'}=-kz^{N+1}\phi_A;\\
    \tilde{c}_z=lz'^{-(N+1)}\tilde{\phi}_A',&\quad\tilde{c}_{z'}=-lz^{-(N+1)}\tilde{\phi}_A.
\end{split}
\end{equation}
Then the result would be:
\begin{equation}\label{cz2}
     2c_{z'}\tilde{c}_{z}=-2\phi_A\tilde{\phi}_A'z^{N+1}z'^{-(N+1)}.
\end{equation}

Note that the equivalence of Eq. \eqref{cz1} and Eq. \eqref{cz2} are guaranteed by Eq. \eqref{eq23}, If we choose the expression Eq. \eqref{cz1}, then we have:
\begin{equation}
\begin{split}
    &2c_{z'}\tilde{c}_{z}\ket{u_{\pm}(z')}\bra{\tilde{u}_{\pm}(z)}\\=&2\begin{pmatrix}
        -\phi_B\tilde{\phi}_B'\phi_A'\tilde{\phi}_A&\mp\phi_B\tilde{\phi}_B'\phi_A'\tilde{\phi}_B\\\mp\phi_B\tilde{\phi}_B'\phi_B'\tilde{\phi}_A&-\phi_B\tilde{\phi}_B'\phi_B'\tilde{\phi}_B
    \end{pmatrix}\\=&\frac{1}{2}\begin{pmatrix}
        -q_{AB}(z')q_{BA}(z)&\mp q_{AB}(z')\\\mp q_{BA}(z)&-1
    \end{pmatrix}.
\end{split}    
\end{equation}
Similar to $C_{bulk}$, we can get $C_{bound}$ that correspond to $G_{bound}$:
\begin{equation}
    (C_{bound})_{kj}=\frac{1}{2}\begin{pmatrix}
        -\slashed{q}_{AA}^{k,j}&\slashed{q}_{AB}^{k,j}\\\slashed{q}_{BA}^{k,j}&-\slashed{q}_{BB}^{k,j}
    \end{pmatrix},
\end{equation}
where:
\begin{small}
\begin{equation}\label{B15}
\begin{split}
    \slashed{q}_{AA}^{k,j}=\sum_z\frac{z'^kz^{-j}}{2N_{\varepsilon}}q_{AB}(z')q_{BA}(z),&\quad\slashed{q}_{AB}^{k,j}=\sum_z\frac{z'^kz^{-j}}{2N_{\varepsilon}} q_{AB}(z'),\\ \slashed{q}_{BA}^{k,j}= \sum_z\frac{z'^kz^{-j}}{2N_{\varepsilon}} q_{BA}(z),&\quad\slashed{q}_{BB}^{k,j}=\sum_z\frac{z'^kz^{-j}}{2N_{\varepsilon}}.
\end{split}    
\end{equation}
\end{small}

If we choose the expression Eq. $\eqref{cz2}$, we can get:
\begin{equation}\label{B16}
\begin{split}
     &\slashed{q}_{AA}^{k,j}=\sum_z\frac{z'^{(k-N-1)}z^{(N+1-j)}}{2N_{\varepsilon}},\\&\slashed{q}_{AB}^{k,j}=\sum_z\frac{z'^{(k-N-1)}z^{(N+1-j)}}{2N_{\varepsilon}}q_{AB}(z),\\&\slashed{q}_{BA}^{k,j}=\sum_z\frac{z'^{(k-N-1)}z^{(N+1-j)}}{2N_{\varepsilon}}q_{BA}(z'),\\&\slashed{q}_{BB}^{k,j}=\sum_z\frac{z'^{(k-N-1)}z^{(N+1-j)}}{2N_{\varepsilon}}q_{AB}(z)q_{BA}(z').
\end{split}
\end{equation}

\begin{widetext}
The Eq. \eqref{B15} and Eq. \eqref{B16} are equal for any finite $N$, where:
\begin{equation}
\begin{split}
  N_{\varepsilon}=  \braket{\Tilde{\psi_{\varepsilon}}|\psi_{\varepsilon}}=&N+c_z\tilde{c}_{z'}\braket{\tilde{u}(z')|u(z)}\braket{\Bar{z'}^{-1}|z}+c_{z'}\tilde{c_z}\braket{\tilde{u}(z)|u(z')}\braket{\Bar{z}^{-1}|z'}\\=&N-\phi_B'\tilde{\phi}_B(\phi_A\tilde{\phi}'_A+\phi_B\tilde{\phi}_B')\braket{\Bar{z'}^{-1}|z}-\phi_B\tilde{\phi}_B'(\phi_A'\tilde{\phi}_A+\phi_B'\tilde{\phi}_B)\braket{\Bar{z}^{-1}|z'}\\=&N-\frac{1}{4}\left[(q_1(z)q_2(z')+1)\braket{\Bar{z'}^{-1}|z}+(q_1(z')q_2(z)+1)\braket{\Bar{z}^{-1}|z'}\right].
\end{split}
\end{equation}
\end{widetext}
This equation will only be useful when doing numerical calculations.

We have obtained the integral of $G_{bulk}$ and $G_{bound}$ so far, since the integral path we consider for ground state usually doesn't cover the origin, the integral for $G_{edge}$ is $0$. The correlation matrix $C$ is:
\begin{equation}
    C=C_{bulk}+C_{bound}=\begin{pmatrix}
        Q_{AA}&Q_{AB}\\Q_{BA}&Q_{BB}
    \end{pmatrix},
\end{equation}
where,
\begin{small}
\begin{equation}
\begin{split}
   (Q_{AA})_{kj}=\frac{1}{2}(q_{\delta}^{k-j}-\slashed{q}_{AA}^{k,j}),\quad& (Q_{AB})_{kj}=\frac{1}{2}(-q_{AB}^{k-j}+\slashed{q}_{AB}^{k,j});\\(Q_{BA})_{kj}=\frac{1}{2}(-q_{BA}^{k-j}+\slashed{q}_{BA}^{k,j}),\quad&(Q_{BB})_{kj}=\frac{1}{2}(q_{\delta}^{k-j}-\slashed{q}_{BB}^{k,j}).
   \end{split}
\end{equation}
\end{small}
We recover Eq. \eqref{57} in Sec. \ref{nu}.

We are going to consider the matrix elements for formal expression next. Transform the sum in Eq. \eqref{B19} into integrate formally, we can get:
\begin{equation}\label{B11}
\begin{split}
 &q_{\delta}^{k-j}=\int_{|z|\equiv\sqrt{|r|}}\frac{dz}{2\pi i z}z^{k-j}=\delta_{k,j},\\& q_{AB}^{k-j}=\int_{|z|\equiv\sqrt{|r|}}\frac{dz}{2\pi i z} q_{AB}(z)z^{k-j},\\& q_{BA}^{k-j}=\int_{|z|\equiv\sqrt{|r|}}\frac{dz}{2\pi i z}q_{BA}(z)z^{k-j}.
\end{split}
\end{equation}
The appearance of $\delta_{k,j}$ is the result of using the residue theorem again.

Transform Eq. \eqref{B15} and Eq. \eqref{B16} into integrate formally, we can get:
\begin{align}\label{B21}
     ^{(1)}\slashed{q}_{AA}^{k,j}&=\int_{|z|\equiv \sqrt{|r|}}\frac{dz}{2\pi i z}q_{AB}(z')q_{BA}(z)z'^kz^{-j},\\ ^{(1)}\slashed{q}_{AB}^{k,j}&=\int_{|z|\equiv \sqrt{|r|}}\frac{dz}{2\pi i z}q_{AB}(z')z'^kz^{-j},\\^{(1)}\slashed{q}_{BA}^{k,j}&=\int_{|z|\equiv \sqrt{|r|}}\frac{dz}{2\pi i z}q_{BA}(z)z'^kz^{-j},\\^{(1)}\slashed{q}_{BB}^{k,j}&=\int_{|z|\equiv \sqrt{|r|}}\frac{dz}{2\pi i z}z'^kz^{-j}.
\end{align}
for Eq. \eqref{B15}.
\begin{small}
\begin{align}
     ^{(2)}\slashed{q}_{AA}^{k,j}&=\int_{|z|\equiv \sqrt{|r|}}\frac{dz}{2\pi i z}z'^{(k-N-1)}z^{(N+1-j)},\\ ^{(2)}\slashed{q}_{AB}^{k,j}&=\int_{|z|\equiv \sqrt{|r|}}\frac{dz}{2\pi i z}q_{AB}(z)z'^{(k-N-1)}z^{(N+1-j)},\\^{(2)}\slashed{q}_{BA}^{k,j}&=\int_{|z|\equiv \sqrt{|r|}}\frac{dz}{2\pi i z}q_{BA}(z')z'^{(k-N-1)}z^{(N+1-j)},\\^{(2)}\slashed{q}_{BB}^{k,j}&=\int_{|z|\equiv \sqrt{|r|}}\frac{dz}{2\pi i z}q_{AB}(z)q_{BA}(z')z'^{(k-N-1)}z^{(N+1-j)}.\label{B28}
\end{align}
\end{small}
for Eq. \eqref{B16}.

The equations with superscript $^{(1)}$ and equations with superscript $^{(2)}$ are generally not equal, that's another mark to the claim that $G_{bound}$ is not well defined in the limit case. However, we believe they both represent the behavior of the system in some extant, so in our validation in Sec. \ref{nu}, we simply choose the expression:
\begin{equation}\label{Blast}
\begin{split}
    \slashed{q}_{AA}^{k,j}=^{(1)}\slashed{q}_{AA}^{k,j}+^{(2)}\slashed{q}_{AA}^{k,j},\\ \slashed{q}_{AB}^{k,j}=^{(1)}\slashed{q}_{AB}^{k,j}+^{(2)}\slashed{q}_{AB}^{k,j},\\
    \slashed{q}_{BA}^{k,j}=^{(1)}\slashed{q}_{BA}^{k,j}+^{(2)}\slashed{q}_{BA}^{k,j},\\ \slashed{q}_{BB}^{k,j}=^{(1)}\slashed{q}_{BB}^{k,j}+^{(2)}\slashed{q}_{BB}^{k,j}.
\end{split}    
\end{equation}

They turn out to correspond with the direct simulation perfectly (see fig. \ref{fig9} and fig. \ref{fig10}).
\bibliography{bib}
\end{document}